\documentclass[11pt,citesort]{article}
\usepackage[latin1]{inputenc}
\usepackage{epsfig}
\usepackage{amsfonts}
\usepackage{amsmath}
\usepackage{amssymb}
\usepackage{xcolor}
\usepackage{cite}
\usepackage{bbold}
\usepackage{hyperref}
\newcommand{\al}{\alpha}
\newcommand{\ksq}{\ka^{\prime 2}}
\newcommand{\bt}{\beta}
\newcommand{\ka}{\kappa}
\newcommand{\dt}{\delta}
\newcommand{\la}{\lambda}
\newcommand{\Or}{\mathcal O}
\newcommand{\vL}{\ensuremath{\mathcal{L}}}

\newcommand{\sq}{^{2}}
\newcommand{\ga}{\gamma}
\newcommand{\GA}{\Gamma}
\newcommand{\TG}{\tilde\Gamma}
\newcommand{\tr}{\mathrm{Tr}}
\newcommand{\bea}{\begin{eqnarray}}
\newcommand{\eea}{\end{eqnarray}}
\newcommand{\bma}{\begin{pmatrix}}
\newcommand{\ema}{\end{pmatrix}}
\newcommand{\nn}{\nonumber}

\begin{document}
\begin{titlepage}

\vspace{2.0cm}

\begin{center}
{\Large\bf Viability of minimal left-right models with discrete symmetries}
\vspace{1.7cm}

{\large \bf Wouter Dekens, Dani\"el Boer} 

\vspace{0.5cm}

{\large 
{\it Van Swinderen Institute, University of Groningen, Nijenborgh 4, 9747 AG Groningen,
The Netherlands}}

\end{center}

\vspace{1.5cm}

\begin{abstract}
We provide a systematic study of minimal left-right models that are invariant under $P$, $C$, and/or $CP$ transformations. Due to the high amount of symmetry such models are quite predictive in the amount and pattern of $CP$ violation they can produce or accommodate at lower energies. Using current experimental constraints some of the models can already be excluded. For this purpose we provide an overview of the experimental constraints on the different left-right symmetric models, considering bounds from colliders, meson-mixing and low-energy observables, such as beta decay and electric dipole moments. The features of the various Yukawa and Higgs sectors are discussed in detail. In particular, we give the Higgs potentials for each case, discuss the possible vacua and investigate the amount of fine-tuning present in these potentials. It turns out that all left-right models with $P$, $C$, and/or $CP$ symmetry have a high degree of fine-tuning, unless supplemented with mechanisms to suppress certain parameters. 
The models that are symmetric under both $P$ and $C$ are not in accordance with present observations, whereas the models with either $P$, $C$, or $CP$ symmetry can not be excluded by data yet. To further constrain and discriminate between the models measurements of $B$-meson observables at LHCb and $B$-factories will be especially important, while measurements of the EDMs of light nuclei in particular could provide complementary tests of the LRMs.
\end{abstract}

\vfill
\end{titlepage}

\section{Introduction}
Left-right models (LRMs) have been studied extensively as possible physics beyond the SM (BSM) \cite{LRSM,Mohapatra:1974hk,Senjanovic1975,Senjanovic1979, Deshpande:1990ip}. LRMs extend the standard model (SM) gauge-group to 
$SU(3)_c\times SU(2)_L\times SU(2)_R\times U(1)_{B-L}$ and possess several attractive features. They offer an interpretation of the $U(1)$ generator in terms of baryon and lepton number and naturally allow for neutrino masses through the see-saw mechanism. Furthermore, the gauge group of the LRM can appear in grand unified theories (GUTs), such as $SO(10)$ and $E_6$, as an intermediate step \cite{Rizzo:1981su}, while avoiding the $SU(5)$ group which has problems with proton decay. But perhaps their most appealing feature is the possibility of having a symmetry between left- and right-handed particles at high energies, a so-called LR symmetry. LR models employing such a symmetry, LR symmetric models (LRSMs), restore parity ($P$) and/or charge conjugation ($C$) invariance at high energies, thereby explaining the $P$- and/or $C$-violating nature of the SM as a low-energy effect. 

From a theoretical standpoint, the most attractive LRSMs might be those which exhibit both $P$ and $C$ symmetries, and thereby $CP$ symmetry, at high energies. Such models {\it in principle} can explain the observed $CP$ violation as resulting from spontaneous $CP$ violation rather than from 
explicit $CP$ violation as in the SM. However, the LR symmetries of such ``$C+P$'' models strongly constrain the left and right CKM matrices, dictating the amount and pattern of $CP$ and flavor violation. For the so-called minimal LRSMs, which are most commonly considered and which have a minimally extended Higgs sector, these model constraints turn out to be incompatible with measurements of Kaon and $B$-meson mixing, as will be discussed. Therefore, minimal LRSMs require explicit $P$ or $C$ violation. It is the goal of this paper to assess the viability of these options, of which many aspects have already been discussed in the literature before. Nevertheless, it seems useful to collect the available results, combine and supplement them, and arrive at clear conclusions about which models are ruled out by current experimental constraints and which models require an unacceptably large amount of fine-tuning. Apart from the LRSMs with ``$C+P$'', $P$ or $C$ symmetry, we also will consider a LRM that is $CP$ symmetric, but not necessarily $P$ and $C$ symmetric. Since this option does not correspond to a LR symmetry, it is not a left-right {\it symmetric} model. 

For all these models we consider the quark and Higgs sectors, review the relations between the left- and right-handed CKM matrices, consider the possible vacua and calculate a measure of the fine-tuning in the Higgs potential in each case. Furthermore, we give an overview of the relevant experimental constraints on the different LRMs, considering bounds from direct searches at the LHC, from $B$-meson-mixing measurements at LHCb and $B$-factories, from Kaon mixing, and low-energy observables, such as beta decay and electric dipole moments (EDMs). As said, in the ``$C+P$'' models current constraints are sufficiently strong to exclude them, but for the other options future measurements, in particular on $CP$ violation by LHCb 
will be able to limit the options further considerably and may also be able to differentiate between the $C$-symmetric and $P$-symmetric LRMs. Measurements on EDMs for the neutron, but also for the proton, other light nuclei and the electron would offer additional tests of LRMs. Currently the LR scale as given by the mass of the right-handed $W$ boson, commonly referred to as $W^\prime$ boson, is required to be at least 2 TeV by direct searches and in the case of $P$ or $C$-symmetric LRSMs 3 TeV by indirect Kaon and $B$-meson constraints \cite{Bertolini:2014sua}. In the coming decade this bound could extend to 8 TeV or higher. As this scale gets pushed upwards, the already considerable if not huge fine-tuning required in the models will increase further and the models become increasingly less likely scenarios. These bounds and perhaps the fine-tuning may be weakened though by considering non-minimal \cite{Ma:1986we,Ashry:2013loa} and/or less symmetric models \cite{Langacker:1989xa,Frank:2010qv,Dev:2013oxa}. We will not include such models here, not for lack of theoretical motivation, but simply in order to limit the scope. 

The large amount of fine tuning in the LRMs models considered here is due to the fact that the Higgs potential necessarily relates the electroweak scale to the LR scale. As the LR scale has been pushed into the TeV range there is a hierarchy between the scales which requires the tuning of some of the parameters in the potential. In fact, unless some of the parameters are chosen to be zero (or exactly related) the fine-tuning becomes extreme. 
Although it is not clear what amount of fine-tuning should be considered acceptable, it does affect the attractiveness of the LRSMs.
We have introduced a measure of fine-tuning often employed in studies of supersymmetric extensions of the SM, in order to quantify the amount of fine-tuning. This may expedite the discussion about the viability of such models and hopefully stimulate the search for new mechanisms to mitigate the fine-tuning problem. 

The outline of this paper is as follows. First we introduce the general minimal LR model in section \ref{introLR} and experimental bounds on $CP$ violation in section \ref{experiments}, while discussing the specific LRSMs in subsequent sections. We first discuss the ``$C+P$'' LRSMs in detail in section \ref{C&Psym}, which, although they turn out not to be viable, have many features in common with the LRSMs with a single LR symmetry to be discussed in section \ref{CorPsym}. We present a summary and conclusions in section \ref{conclusion}.

\section{Minimal left-right models}\label{introLR}
In this section we will discuss minimal left-right models and highlight some of their features. We will start with the basic ingredients of the model, namely, its field content.
\subsection{Field content}
The gauge group of left-right (LR) models is given by $SU(2)_L \times SU(2)_R \times U(1)_{B-L}$ \cite{LRSM,Mohapatra:1974hk,Senjanovic1975,Senjanovic1979, Deshpande:1990ip}. As in the standard model (SM) the left-handed fermions form doublets under $SU(2)_L$. New, with respect to the SM, is that the right-handed fermions now form doublets under the added gauge group,  $SU(2)_R$. In order to build these doublets right-handed neutrinos have to be introduced. In short, the fermions are assigned to representations of the above gauge group as follows,
\bea Q_L &=&\bma u_L\\d_L\ema \in (2,1,1/3), \qquad Q_R = \bma u_R\\d_R\ema\in (1,2,1/3),\nn\\
L_L &=&\bma \nu_L\\l_L\ema \in (2,1,-1), \qquad L_R = \bma \nu_R\\l_R\ema\in (1,2,-1).\eea
With the fermions in the above representations a scalar, $\phi\in (2,2^*,0)$, is required in order to produce fermion masses.  
Furthermore, additional scalar fields are introduced to facilitate the breakdown of the LR gauge group to that of the SM. In the LR model under discussion here, which has been often considered e.g.\ \cite{Deshpande:1990ip, Zhang2008}, these are two triplets $\Delta_{L,R}$ assigned to $(3,1,2)$ and $(1,3,2)$, respectively. These fields can be written as
\bea \phi = \bma \phi_1^0 & \phi_1^+\\ \phi_2^- & \phi_2^0 \ema ,\qquad
\Delta_{L,R} = \bma \delta^+_{L,R}/\sqrt{2} & \delta^{++}_{L,R} \\ \delta^0_{L,R} & -\delta^+_{L,R}/\sqrt{2} \ema .
\label{scalars}\eea
We will refer to LR models with such a Higgs sector as \textit{minimal} LRMs. Symmetry breaking is realized through the vacuum expectation values (vevs) of the scalar fields, 
\bea
\langle \phi \rangle =\sqrt{1/2} \bma \kappa &0\\0&\kappa' e^{i\al} \ema ,\qquad \langle \Delta_{L} \rangle = \sqrt{1/2}\bma 0&0\\v_{L}e^{i\theta_L}&0\ema ,\qquad \langle \Delta_{R} \rangle = \sqrt{1/2}\bma 0&0\\v_{R}&0\ema ,
\eea
where all parameters are real after gauge transformations have been used to eliminate two of the possible phases \cite{Deshpande:1990ip}. In the first step of symmetry breaking the vev of the right-handed triplet, $v_R$, breaks the $SU(2)_L \times SU(2)_R \times U(1)_{B-L}$ group down to $SU(2)_L \times U(1)_{Y}$. This vev also defines the high scale of the model, and gives the main contribution to the masses of the additional gauge bosons, $W_R^\pm$ and $Z_R$ belonging to $SU(2)_R$. At the electroweak scale the vevs of the bidoublet, $\ka$ and $\ka'e^{i\al}$, then break $SU(2)_L \times U(1)_{Y}$ to $U(1)_{\text{EM}}$. In turn, these vevs dictate the masses of the $W_L^\pm$ and $Z_L$ bosons, which belong to $SU(2)_L$, while $\al$ is the parameter indicating spontaneous $CP$ violation. This implies these vevs are of the electroweak scale, indeed, we have
\bea \ka_+=v\simeq 246 \, \text{GeV}, \eea
where $\ka_{\pm}\sq \equiv \kappa\sq \pm\kappa^{\prime 2}$. Finally, while the Dirac masses of the fermions are generated by the vevs of $\phi$, the vevs of the triplets generate Majorana masses for the neutrinos. Thus, $v_L$ contributes to the light neutrino masses and so is not expected to exceed this scale by much, i.e., $v_L \lesssim \Or(1 \,\text{eV})$. A (much) less stringent upper bound without theoretical prejudice can be derived from the $\rho_0$ parameter; $\rho_0\equiv \frac{M_W\sq}{M_Z\sq c_W\sq}$ defined such that it is $1$ to all orders in the SM. Since $v_L$ breaks custodial symmetry it contributes to $\rho_0-1$, from a global fit of $\rho_0$ \cite{pdg:2012} one can then deduce, $v_L\lesssim 5\,\text{GeV}$.

As for the Higgs fields themselves, there are two neutral and two singly-charged would-be-Goldstone bosons. The remaining fields are physical and make up six neutral, two singly-charged and two doubly charged fields. For approximate expressions of the mass eigenstates and their masses, see e.g. \cite{Zhang2008,Duka:1999uc,Kiers:2005gh}. One of the mass eigenstates plays the part of a SM-like scalar whose mass should be $125 \, \text{GeV}$. The non-SM neutral fields arising from the bidoublets are required to be heavy, $>10$ TeV in LRMs \cite{Ball:1999mb, Bertolini:2014sua}, as they give rise to stringently constrained flavor-changing neutral currents, see section \ref{CP1Higgs}. The scalars arising from the triplet fields are not as well constrained and can still be relatively light while keeping the flavor-changing scalars heavy \cite{Bambhaniya:2013wza}. In fact, the doubly charged scalars can still have masses $\sim 450 \,\text{GeV}$, while in the future, at $\sqrt{s}=14$ TeV with $300$ fb$^{-1}$, the LHC is expected to probe masses up to $600$ GeV
\cite{Bambhaniya:2013wza,Bambhaniya:2014cia}.

\subsection{LR symmetries}
One of the main motivations for LRMs is the possibility of explaining the broken symmetry between left and right in the SM as a low-energy phenomenon. In  LR models it is possible to restore this symmetry at high energies, which is then spontaneously broken at lower energies by the vevs of the scalar fields. There are two possible transformations which qualify as symmetries between left and right\footnote{There are two other possible transformations on the $\phi$ fields which would qualify as a left-right symmetry, namely, $\phi\rightarrow\tilde\phi^\dagger$ and $\phi\rightarrow\tilde\phi^T$, where $\tilde\phi = \tau_2 \phi^*\tau_2$. However, as observed in \cite{Maiezza:2010ic} these lead to unrealistic quark mass matrices, $M_u=M_d^\dagger$ in the former case and $\tr(M_uM_u^\dagger)=\tr(M_dM_d^\dagger)$ in the latter.}\footnote{
To be general these transformations should also include the possibility of changing the flavors of the quarks. However, when a single LR symmetry applies we can always choose a basis such that the transformations are as in Eq.\ \eqref{C&Ptransf}. When both LR symmetries apply flavor rotations can play a role as we will see in section \ref{C&Psym}.}
\bea
&P:& \qquad Q_{L}\longleftrightarrow Q_R , \,\quad \qquad\phi \longleftrightarrow \phi^{\dagger},\qquad \Delta_{L,R}\longleftrightarrow \Delta_{R,L},\nn\\
&C:&\qquad Q_{L}\longleftrightarrow (Q_R)^c , \qquad\phi\longleftrightarrow \phi^{T} ,\qquad \Delta_{L,R}\longleftrightarrow \Delta_{R,L}^*,
\label{C&Ptransf}\eea
where the superscript $c$ indicates charge conjugation. A LR model with such a $P$ or $C$ symmetry is called left-right symmetric. Note that the combination of the two symmetries in Eq.~\eqref{C&Ptransf}, $CP$, does not interchange left- and right-handed fields and so is not a LR symmetry. Both LR symmetries require the $SU(2)_{L,R}$ gauge couplings to be equal, $g_L=g_R$, at the LR scale, although a difference between the two could be induced when they are evolved down to the electroweak scale. Two more specific (albeit not necessarily minimal) models, often discussed in the literature, are the manifest and pseudomanifest LR models. The former refers to a LR model with $P$-symmetric Yukawa couplings and the additional assumption of a vanishing spontaneous phase, $\al=0$ \cite{Beg:1977ti,Langacker:1989xa}. A pseudomanifest LR model on the other hand assumes $C$- and $P$-symmetric Yukawa couplings \cite{Harari:1983gq,Langacker:1989xa}. In the past the case of a $P$-symmetric LR model was mainly studied \cite{LRSM,Mohapatra:1974hk,Zhang2008}, however, recently there has been renewed interest in the $C$-symmetric case as well \cite{Maiezza:2010ic, Bertolini:2014sua}. In either case these symmetries impose important restrictions, as we will see later.

\subsection{Charged gauge-bosons}
Perhaps the most characteristic way in which LR models affect observables is through the right-handed charged-current interaction of the $W_R^\pm$ boson. For the quarks it is given by (in the quark-mass basis)
\bea
\vL^{\text{CC}} = \frac{g_L}{\sqrt{2}}\overline U_L \gamma^\mu V_L D_L W_{L\mu}^+ + \frac{g_R}{\sqrt{2}}\overline U_R \gamma^\mu V_R D_R W_{R\mu}^+ +\text{h.c.}\,,
\eea
where $V_{L}$ and $V_R$ are the SM CKM matrix and its right-handed equivalent. However, the gauge fields $W_{L,R}^\pm$ are not quite mass eigenstates. The two charged gauge-bosons mix because both of them couple to the bidoublet $\phi$ which is charged under both $SU(2)$ groups. The mass terms for the charged gauge-bosons are given by
\bea
\vL^{W_{\text{mass}}} = (W^-_{L\mu}\quad W^-_{R\mu}) \bma \frac{g_L\sq}{4}(\kappa\sq+\kappa^{\prime 2}+2v_L\sq)& -\frac{1}{2}g_L g_R \kappa \kappa'e^{-i\al} \\ -\frac{1}{2}g_L g_R \kappa \kappa' e^{i\al}& \frac{g_R\sq}{4}(\kappa\sq+\kappa^{\prime 2}+2v_R\sq)\ema
\bma W^{+\mu}_L\\ W^{+\mu}_R \ema ,
\eea
where $g_{L,R}$ are the coupling constants of $SU(2)_{L,R}$. These gauge couplings will be equal in both the $P$- and $C$-symmetric case. The gauge eigenstates are related to the mass eigenstates as follows
\bea
\bma W^{+\mu}_L\\ W^{+\mu}_R \ema = \bma \cos\zeta &-\sin\zeta e^{-i\al}\\ \sin\zeta e^{i\al}&\cos\zeta\ema \bma W^{+\mu}_1\\ W^{+\mu}_2 \ema ,\qquad 
\tan\zeta \simeq \frac{g_L}{g_R}\frac{\kappa\kappa'}{v_R\sq},
\eea
where $W_{1,2}^\pm$ refer to the mass eigenstates of the charged gauge-bosons. The masses themselves are approximately given by
\bea
M_1\sq\simeq \frac{g_L\sq\ka_+\sq}{4},\qquad M_{2}\sq\simeq \frac{g_R\sq v_R\sq}{2}.
\eea

Direct searches at the LHC set a lower limit of 2 TeV (95\% CL) on the mass of the right-handed $W_R^\pm$ from the $W_R^+\rightarrow t\bar b$ channel \cite{Chatrchyan:2014koa}. More stringent limits have been obtained in leptonic decays which rely on certain assumptions about right-handed neutrinos. These limits extend to $M_2\geq 2.5-3$ TeV \cite{CMS:2012zv,ATLAS:2012ak,Khachatryan:2014dka} for a range of values for the right-handed neutrino mass, $M_N$. Although these bounds on $M_2$ depend on $M_N$ the two masses become correlated in some LRSMs after applying constraints from low-energy precision experiments in muon decay, thereby considerably reducing the allowed region in parameter space \cite{Chakrabortty:2012pp}.
The collider bounds from both types of channels assume the right-handed couplings to be the same as the left-handed couplings, e.g.\ for the $W_R^+\rightarrow t\bar b$ channel $g_L |V_L^{tb}|= g_R |V_R^{tb}|$. Thus, the strength of the above bounds is in part determined by whether or not the model is  LR symmetric. If we do not assume any LR symmetry these bounds can be weakened and even be evaded in some cases.

\subsection{Yukawa couplings}
In turn, the masses for the quarks are generated by the interactions of the bidoublet with the quarks. The most general form of the Yukawa interactions respecting the gauge symmetries in the weak basis is,
\bea 
-\vL_Y = \bar Q_L\big( \GA \phi + \TG \tilde \phi \big)Q_R +\text{h.c.}\quad ,
\label{Yukawa}
\eea
where $\Gamma$ and $\TG$ are complex $3\times 3$ matrices and $\tilde \phi \equiv \tau_2 \phi^* \tau_2$. After the Higgs fields acquire their vevs this leads to the following mass matrices for the quarks
\bea
M_u = \sqrt{1/2}(\ka \Gamma +\ka' e^{-i\al}\TG),\qquad M_d = \sqrt{1/2}(\ka'e^{i\al} \Gamma +\ka\TG).\label{masses}
\eea
The implications of the possible LR symmetries on the Yukawa sector are the following
\bea
P: \qquad \Gamma &=& \Gamma^\dagger,\qquad \TG = \TG^\dagger,\\
 C: \qquad \Gamma &=& \Gamma^T,\qquad \TG = \TG^T. \label{implications}
\eea
For the $P$-symmetric case this means that if $\al$ were zero, as in the manifest LR symmetric model, the mass matrices would be hermitian as well. In this limit there is a relation between the left- and right-handed CKM matrices, namely, 
\bea V_R = S_u V_L S_d, \label{CKMP}\eea
where $S_{u,d}$ are diagonal matrices of signs.
In general $\al\neq 0$ and the above relation will not be satisfied. Nonetheless, in the $P$-symmetric case, in order to reproduce the observed quark masses, the combination $\ka'/\ka \sin\al$ should be small \cite{Maiezza:2010ic}. Thus, the quark mass matrices will be nearly hermitian, implying that Eq.~\eqref{CKMP} is approximately correct and the right and left mixing angles should be nearly equal. This was already shown numerically in Refs. \cite{Kiers:2002cz,Maiezza:2010ic} and was recently confirmed by an explicit solution of $V_R$ \cite{Senjanovic:2014pva}.

In the $C$-symmetric case the mass matrices will be symmetric which implies the following relation between the two CKM matrices \cite{Branco:1982wp}
\bea V_R = K_u V_L^* K_d, \label{CKMC}\eea
where $K_{u}=\text{diag}(\theta_u,\theta_c,\theta_t)$ and $K_{d}=\text{diag}(\theta_d,\theta_s,\theta_b)$ are diagonal matrices of phases, of which one combination can be set to zero, while the rest remains unconstrained. This relation holds irrespective of the value of $\al$. As a result the mixing angles in both matrices will be equal. 

Which relation between the left- and right-handed CKM matrices applies has implications for the bounds that can be set on these models. We will come back to this issue when discussing the $P$- and $C$-symmetric LR models in more detail.

\subsection{The Higgs potential}
The final part of the Lagrangian to be discussed is the Higgs potential. As we intend to discuss the potential for the LR symmetries, $P$ and $C$ as well as the $CP$-symmetric case, we will give the potentials for these three cases.

The potential invariant under the gauge group and the $P$ symmetry is given by \cite{Deshpande:1990ip} 
\bea
V^P_H&=&
-\mu_1\sq \,\text{Tr}(\phi^\dagger \phi) -\mu_2\sq \big[\text{Tr}(\tilde\phi^\dagger \phi)+\text{Tr}(\phi^\dagger\tilde \phi)\big] - \mu_3\sq \big[\text{Tr}(\Delta_L\Delta_L^\dagger)+\text{Tr}(\Delta_R\Delta_R^\dagger)\big]+\lambda_1 \,\big[\text{Tr}(\phi^\dagger \phi)\big]\sq\nn\\
&&+\la_2\big(\big[\text{Tr}(\tilde\phi^\dagger \phi)\big]\sq+\big[\text{Tr}(\phi^\dagger\tilde \phi)\big]\sq\big)+\la_3 \, \text{Tr}(\tilde\phi^\dagger \phi)\,\text{Tr}(\phi^\dagger\tilde \phi)+\la_4\,\text{Tr}(\phi^\dagger \phi)\big[\text{Tr}(\tilde\phi^\dagger \phi)+\text{Tr}(\phi^\dagger\tilde \phi)\big]\nn\\
&&+\rho_1\big(\big[\text{Tr}(\Delta_L\Delta_L^\dagger)\big]\sq+\big[\text{Tr}(\Delta_R\Delta_R^\dagger)\big]\sq\big)\nn\\
&&+\rho_2 \big[\text{Tr}(\Delta_L\Delta_L)\text{Tr}(\Delta_L^\dagger\Delta_L^\dagger)+\text{Tr}(\Delta_R\Delta_R)\text{Tr}(\Delta_R^\dagger\Delta_R^\dagger)\big]+\rho_3\text{Tr}(\Delta_L\Delta_L^\dagger)\text{Tr}(\Delta_R\Delta_R^\dagger)\nn\\
&&+\rho_4\big[\text{Tr}(\Delta_L\Delta_L)\text{Tr}(\Delta_R^\dagger\Delta_R^\dagger)+\text{Tr}(\Delta_R\Delta_R)\text{Tr}(\Delta_L^\dagger\Delta_L^\dagger)\big]\nn\\
&&+\al_1 \,\text{Tr}(\phi^\dagger \phi)\big[\text{Tr}(\Delta_L\Delta_L^\dagger)+\text{Tr}(\Delta_R\Delta_R^\dagger)\big]\nn\\
&&+\al_2\big(e^{i\delta_2}\big[\text{Tr}(\tilde\phi^\dagger \phi)\text{Tr}(\Delta_R\Delta_R^\dagger)+\text{Tr}(\phi^\dagger\tilde \phi)\text{Tr}(\Delta_L\Delta_L^\dagger)\big]+\text{h.c.}\big)\nn\\
&&+\al_3\big[\text{Tr}(\phi\phi^\dagger \Delta_L\Delta_L^\dagger)+\text{Tr}(\phi^\dagger \phi\Delta_R\Delta_R^\dagger)\big]+\bt_1\big[\text{Tr}(\phi \Delta_R \phi^\dagger\Delta_L^\dagger)+\text{Tr}(\phi^\dagger \Delta_L \phi\Delta_R^\dagger)\big]\nn\\
&&+\bt_2\big[\text{Tr}(\tilde\phi \Delta_R \phi^\dagger\Delta_L^\dagger)+\text{Tr}(\tilde\phi^\dagger \Delta_L \phi\Delta_R^\dagger)\big] +\bt_3\big[\text{Tr}(\phi \Delta_R \tilde\phi^\dagger\Delta_L^\dagger)+\text{Tr}(\phi^\dagger \Delta_L \tilde\phi\Delta_R^\dagger)\big],
\label{HiggspotP}\eea
while the potential in the case of an unbroken $C$ symmetry at high energies is given by,
\bea
V_H^C&=&
-\mu_1\sq \,\text{Tr}(\phi^\dagger \phi) -\mu_2\sq \big[e^{i\delta_{\mu_2}}\text{Tr}(\tilde\phi \phi^\dagger)+\text{h.c.}\big] - \mu_3\sq \big[\text{Tr}(\Delta_L\Delta_L^\dagger)+\text{Tr}(\Delta_R\Delta_R^\dagger)\big]+\lambda_1 \,\big[\text{Tr}(\phi^\dagger \phi)\big]\sq\nn\\
&&+\la_2\big(e^{i\delta_{\la_2}}\big[\text{Tr}(\tilde\phi \phi^\dagger)\big]\sq+\text{h.c.}\big)+\la_3 \, \text{Tr}(\tilde\phi^\dagger \phi)\,\text{Tr}(\phi^\dagger\tilde \phi)+\la_4\,\text{Tr}(\phi^\dagger \phi)\big[e^{i\delta_{\la_4}}\text{Tr}(\tilde\phi \phi^\dagger)+\text{h.c.}\big]\nn\\
&&+\rho_1\big(\big[\text{Tr}(\Delta_L\Delta_L^\dagger)\big]\sq+\big[\text{Tr}(\Delta_R\Delta_R^\dagger)\big]\sq\big)\nn\\
&&+\rho_2 \big[\text{Tr}(\Delta_L\Delta_L)\text{Tr}(\Delta_L^\dagger\Delta_L^\dagger)+\text{Tr}(\Delta_R\Delta_R)\text{Tr}(\Delta_R^\dagger\Delta_R^\dagger)\big]+\rho_3\text{Tr}(\Delta_L\Delta_L^\dagger)\text{Tr}(\Delta_R\Delta_R^\dagger)\nn\\
&&+\rho_4\big[e^{{-i\dt_{\rho_4}}}\text{Tr}(\Delta_L\Delta_L)\text{Tr}(\Delta_R^\dagger\Delta_R^\dagger)+e^{{i\dt_{\rho_4}}}\text{Tr}(\Delta_R\Delta_R)\text{Tr}(\Delta_L^\dagger\Delta_L^\dagger)\big]\nn\\
&&+\al_1 \,\text{Tr}(\phi^\dagger \phi)\big[\text{Tr}(\Delta_L\Delta_L^\dagger)+\text{Tr}(\Delta_R\Delta_R^\dagger)\big]\nn\\
&&+\al_2\big[e^{i\delta_{\al_2}}\text{Tr}(\tilde \phi^\dagger\phi)+\text{h.c.}\big]\big[\text{Tr}(\Delta_L\Delta_L^\dagger)+\text{Tr}(\Delta_R\Delta_R^\dagger)\big]\nn\\
&&+\al_3\big[\text{Tr}(\phi\phi^\dagger \Delta_L\Delta_L^\dagger)+\text{Tr}(\phi^\dagger \phi\Delta_R\Delta_R^\dagger)\big]\nn\\
&&+\bt_1\big[e^{i\dt_{\bt_1}}\text{Tr}(\phi \Delta_R \phi^\dagger\Delta_L^\dagger)+e^{-i\dt_{\bt_1}}\text{Tr}(\phi^\dagger \Delta_L \phi\Delta_R^\dagger)\big]\nn\\
&&+\bt_2\big[e^{i\dt_{\bt_2}}\text{Tr}(\tilde\phi \Delta_R \phi^\dagger\Delta_L^\dagger)+e^{-i\dt_{\bt_2}}\text{Tr}(\tilde\phi^\dagger \Delta_L \phi\Delta_R^\dagger)\big] \nn\\
&&+\bt_3\big[e^{i\dt_{\bt_3}}\text{Tr}(\phi \Delta_R \tilde\phi^\dagger\Delta_L^\dagger)+e^{-i\dt_{\bt_3}}\text{Tr}(\phi^\dagger \Delta_L \tilde\phi\Delta_R^\dagger)\big].
\label{HiggspotC}\eea
Finally, the $CP$-symmetric, but not necessarily $C$- or $P$-symmetric, potential is given by,
\bea
V^{CP}_H&=&
-\mu_1\sq \,\text{Tr}(\phi^\dagger \phi) -\mu_2\sq \big[\text{Tr}(\tilde\phi^\dagger \phi)+\text{Tr}(\phi^\dagger\tilde \phi)\big] - \mu_{3L}\sq \text{Tr}(\Delta_L\Delta_L^\dagger)-\mu_{3R}\sq\text{Tr}(\Delta_R\Delta_R^\dagger)\nn\\
&&+\lambda_1 \,\big[\text{Tr}(\phi^\dagger \phi)\big]\sq+\la_2\big(\big[\text{Tr}(\tilde\phi^\dagger \phi)\big]\sq+\big[\text{Tr}(\phi^\dagger\tilde \phi)\big]\sq\big)\nn\\
&&+\la_3 \, \text{Tr}(\tilde\phi^\dagger \phi)\,\text{Tr}(\phi^\dagger\tilde \phi)+\la_4\,\text{Tr}(\phi^\dagger \phi)\big[\text{Tr}(\tilde\phi^\dagger \phi)+\text{Tr}(\phi^\dagger\tilde \phi)\big]\nn\\
&&+\rho_{1L}\big[\text{Tr}(\Delta_L\Delta_L^\dagger)\big]\sq+\rho_{1R}\big[\text{Tr}(\Delta_R\Delta_R^\dagger)\big]\sq\nn\\
&&+\rho_{2L} \text{Tr}(\Delta_L\Delta_L)\text{Tr}(\Delta_L^\dagger\Delta_L^\dagger)+\rho_{2R}\text{Tr}(\Delta_R\Delta_R)\text{Tr}(\Delta_R^\dagger\Delta_R^\dagger)\nn\\
&&+\rho_3\text{Tr}(\Delta_L\Delta_L^\dagger)\text{Tr}(\Delta_R\Delta_R^\dagger)\nn\\
&&+\rho_4\big[\text{Tr}(\Delta_L\Delta_L)\text{Tr}(\Delta_R^\dagger\Delta_R^\dagger)+\text{Tr}(\Delta_R\Delta_R)\text{Tr}(\Delta_L^\dagger\Delta_L^\dagger)\big]\nn\\
&&+\al_{1L} \,\text{Tr}(\phi^\dagger \phi)\text{Tr}(\Delta_L\Delta_L^\dagger)+\al_{1R}\,\text{Tr}(\phi^\dagger \phi)\text{Tr}(\Delta_R\Delta_R^\dagger)\nn\\
&&+\big[\text{Tr}(\tilde\phi^\dagger \phi)+\text{Tr}(\phi^\dagger\tilde \phi)\big]\big[\al_{2L}\text{Tr}(\Delta_L\Delta_L^\dagger)+\al_{2R}\text{Tr}(\Delta_R\Delta_R^\dagger)\big]\nn\\
&&+\al_{3L}\text{Tr}(\phi\phi^\dagger \Delta_L\Delta_L^\dagger)+\al_{3R}\text{Tr}(\phi^\dagger \phi\Delta_R\Delta_R^\dagger)+\bt_1\big[\text{Tr}(\phi \Delta_R \phi^\dagger\Delta_L^\dagger)+\text{Tr}(\phi^\dagger \Delta_L \phi\Delta_R^\dagger)\big]\nn\\
&&+\bt_2\big[\text{Tr}(\tilde\phi \Delta_R \phi^\dagger\Delta_L^\dagger)+\text{Tr}(\tilde\phi^\dagger \Delta_L \phi\Delta_R^\dagger)\big] +\bt_3\big[\text{Tr}(\phi \Delta_R \tilde\phi^\dagger\Delta_L^\dagger)+\text{Tr}(\phi^\dagger \Delta_L \tilde\phi\Delta_R^\dagger)\big].
\label{HiggspotCP}\eea
In all cases all parameters are real, the $P$-symmetric potential contains 18 parameters while in the $C$- and $CP$-symmetric cases there are 25 and 23 parameters, respectively. This implies that the $P$-symmetry is the most constraining when it comes to the Higgs potential. However, as we will see later these potentials are all closely related.

In the upcoming discussion about the amount of fine-tuning in these potentials we impose the condition that the dimensionless parameters are in the perturbative regime, i.e.\ take on values of order 1. As will be discussed, in some cases the amount of fine-tuning can be greatly reduced by setting some parameters to zero. In the literature this has been done for the $\beta_i$ parameters, without further justification, e.g.\ \cite{Zhang2008}. This in turn requires $v_L=0$. 
The choice $\bt_i=0$ is in principle unstable under renormalization unless enforced by a symmetry. However, in Refs.~\cite{Deshpande:1990ip,Barenboim:2001vu} it was argued that such symmetries do not allow for Majorana masses for the neutrinos, it may thus not be a viable option. 

We note that for avoiding or strongly reducing the fine-tuning it is not needed to set $\bt_i=0$. As we will demonstrate below, the same reduction in the amount of fine-tuning can be achieved by arranging $v_L=0$, which does however require relations among some of the $\beta_i$ parameters \cite{Barenboim:2001vu}. Another option is to introduce a mechanism that yields small $\beta_i$. In Ref.\ \cite{Kiers:2005gh} a softly broken horizontal $U(1)$ symmetry was introduced to enforce $\beta_i$ of order $v_L/v_R$, with $v_L \sim 0.1\ {\rm eV}$. This also reduces the fine-tuning and satisfies the current experimental constraints.  

\section{Experimental constraints on $CP$ violation}\label{experiments}
In this section we will discuss a number of experimental constraints on $CP$ violation in LRMs, namely those from Kaon mixing and decays, $\overline B_{d,s}-B_{d,s}$ mixing, electric dipole moments (EDMs) and neutron $\bt$ decay. We will discuss the impact of these bounds in specific LRSMs in more detail in the subsequent sections.

\subsection{Kaon mixing and decays}

The well-known indirect and direct $CP$-violating parameters in the Kaon sector, $\varepsilon$ and $\varepsilon^{\prime}$, are currently determined to be \cite{pdg:2012}
\begin{eqnarray}
|\varepsilon| & = & (2.228 \pm 0.011) \cdot 10^{-3},\\ 
{\text Re}(\varepsilon^\prime/\varepsilon) & = & (1.65 \pm 0.26) \cdot 10^{-3}.
\end{eqnarray}
Depending on the particular realization of the LR symmetry, these parameters can lead to strong constraints on the phases in the matrix $K_{u,d}$ relating the left and right CKM matrices. For the relevant expressions we refer to Refs.\ \cite{Ecker:1985vv,Basecq:1985cr,Bertolini:2012pu,Bertolini:2014sua}.

In LRMs there are additional contributions to $\varepsilon$ compared to the SM from box diagrams involving $W_R^{\pm}$ bosons and tree-level diagrams involving flavor-changing Higgs bosons \cite{Ecker:1985ei}. This is analogous to the case of $B$ meson mixing which will be discussed more explicitly next. 

\subsection{$\overline B_{d,s}-B_{d,s}$ mixing}
\begin{figure}[t!]
\centering
\includegraphics[width=100mm]{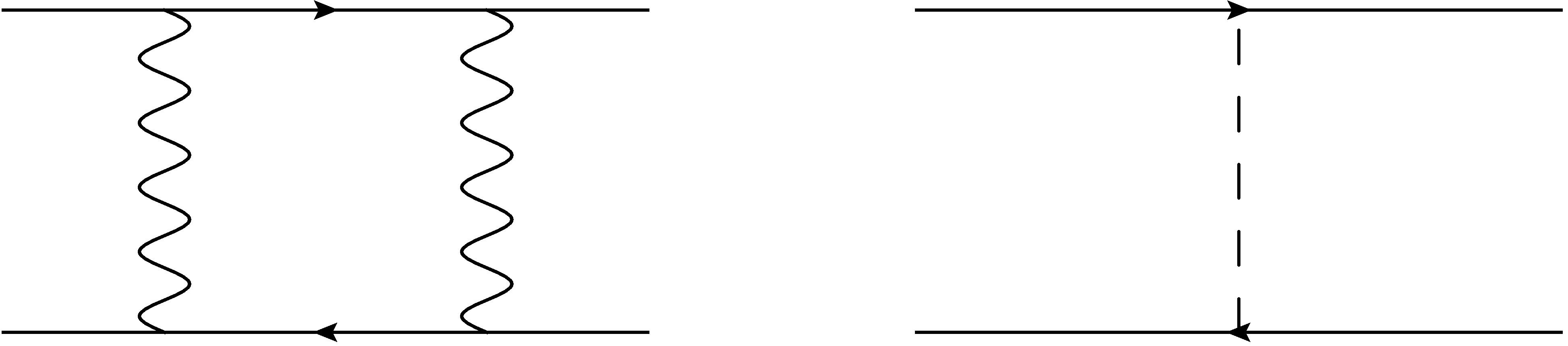} \qquad \qquad 
\includegraphics[width=42mm]{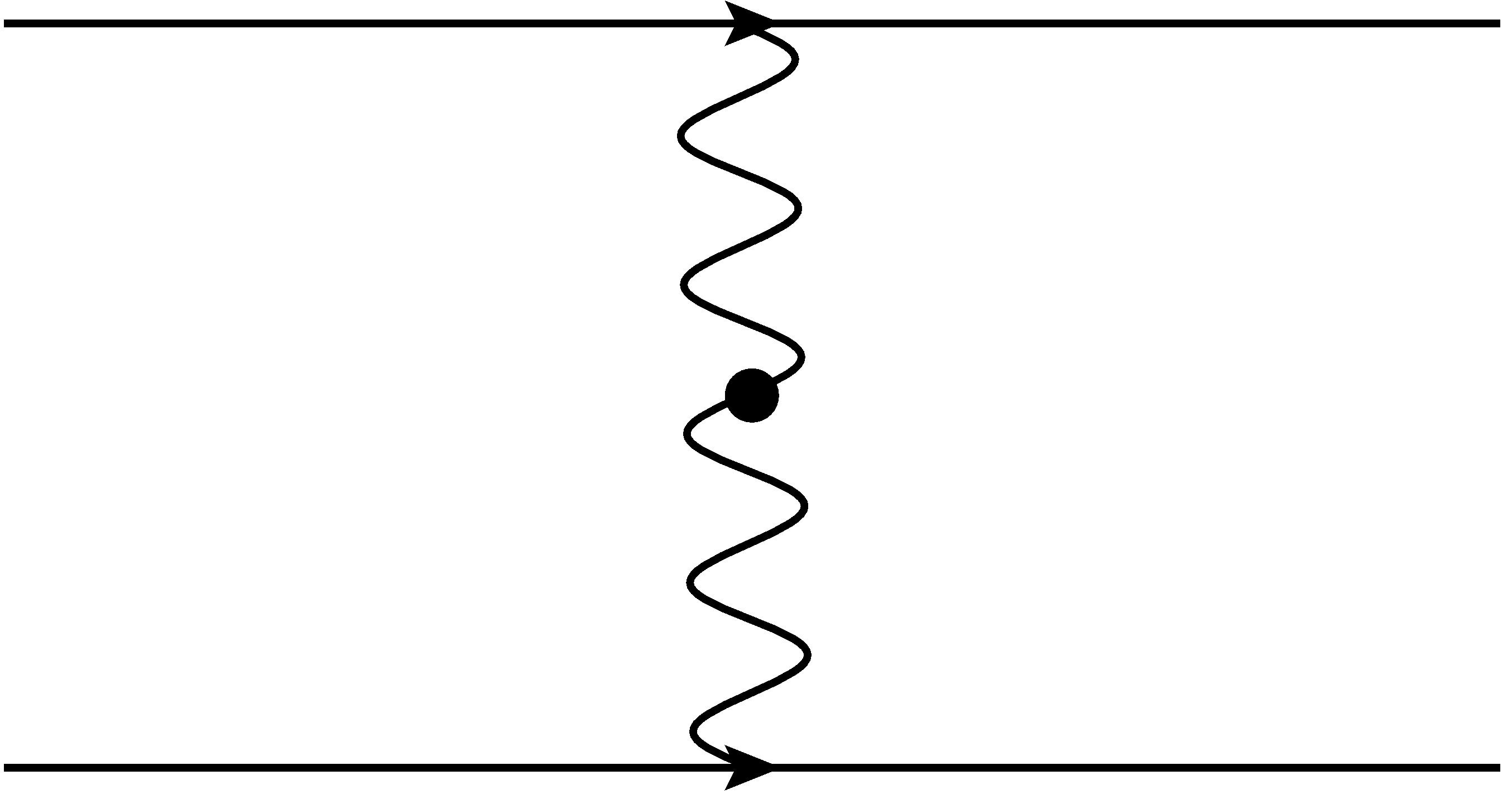} \\
a \hspace*{5.4cm} b \hspace*{5.4cm} c
\caption{Figures a and b show some of the LR contributions to meson mixing. The wavy and dashed lines represent $W_{L,R}^\pm$-bosons and flavor-changing Higgs bosons, respectively. The external fermions lines are the quarks in the mesons. Figure c shows the dominant diagram contributing to the neutron EDM and $CP$ violation in neutron $\bt$ decay (assuming $\bar\theta = 0$). The fermion lines now represent up and down quarks while one line represents $e$ and $\nu$ in case of $\bt$ decay and the dot denotes $W_L$-$W_R$ mixing.
} 
\label{diagrams}
\end{figure} 
The $\overline B_{d,s}-B_{d,s}$ mixing is described by the off-diagonal matrix element $M_{12}^q= \langle B_q|\mathcal H |\overline  B_q\rangle/2M_{B_q}$. In the SM $M_{12}^q$ is determined by box diagrams involving $W_L^\pm$ bosons. The magnitude of $M_{12}^q$ is related to the mass difference $\Delta M_{B_q}$ between the mesons while its phase signifies $CP$ violation,
\bea
\Delta M_{B_q} = 2|M_{12}^q|,\qquad \phi_q = \text{Arg} M_{12}^q. 
\eea
In LRMs there are additional contributions from box diagrams involving $W_R^{\pm}$ bosons and tree-level diagrams involving flavor-changing Higgs bosons. Separating the SM and LRM contributions, $M_{12} = M_{12}^\text{{SM}} + M_{12}^{\text{LR}}$ the new contributions can be parametrized by the following quantities,
\bea 
M_{12}^q = M_{12}^{\text{SM}}(1+h_q),\qquad h_q \equiv \frac{M_{12}^{\text{LR}}}{ M_{12}^\text{{SM}}}, \qquad h_q = |h_q| e^{i\sigma_q}. 
\eea
Thus, the magnitude of $1+h_q$ can be constrained by the mass differences, while $CP$ violation in $B_q$ mixing as measured by $\phi_{d,s}$ is sensitive to its phase. The LR contribution to these angles is given by
\bea \phi_q^{\text{LR}} = \text{Arg} (1+|h_q| e^{i\sigma_q}),\qquad 
 \sigma_q \simeq \text{Arg}\bigg(-\frac{V_R^{tb}V_R^{tq\,*}}{V_L^{tb}V_L^{tq\,*}}\bigg). \label{phis}
\eea
The expressions for $h_q$ can be found in 
Refs.~\cite{Ecker:1985vv,Basecq:1985cr,Bertolini:2014sua}. Clearly, this contribution depends on the phases present in the CKM matrices. This in turn depends on the choice of LR symmetry. When the phases in $V_R$ are free they can be tuned so as to avoid the bounds from the $CP$-violating observables, $\phi_{d,s}$. However, in the more constrained LRSMs these bounds will be important. These phases appear in asymmetries of $B_{d,s}$ decays, currently the averages of experimental measurements give the following values \cite{HFAG} 
\bea
-\mathcal{A}_{CP}^{\text{mix}}(B_d\rightarrow f) &=& \sin\phi_d = 0.68 \pm 0.02\quad (68\%\,\text{CL}),\label{phid} \\
\mathcal{A}_{CP}^{\text{mix}}(B_s\rightarrow f')&=&\sin\phi_s = 0.00\pm 0.07 \quad (68\%\,\text{CL}),\eea  
where $f = (J/\psi \,K_S,\,J/\psi \,K_L,\dots)$ and $f' = (J/\psi \,\phi,\, J/\psi\, f_0(980),\dots)$ are all final states involving $\bar cc\bar s d$  and $\bar cc\bar ss$ valence quarks, respectively. Currently the value for $\sin\phi_d$ is still compatible with its SM prediction, $\sin\phi_d^{\text{SM}} \sim 0.83$ \cite{Charles:2011va}, within (theoretical) errors, but is approaching a 3$\sigma$ level deviation \cite{Lenz:2012az}. The SM prediction for $\sin\phi_s^{\text{SM}} \sim 0.036$ is also still consistent with the experimental value \cite{Charles:2011va}. The precision of these measurements is expected to improve of course. In the long run, the error in the $\phi_{d}$ ($\phi_{s}$) measurements should decrease by roughly a factor 3 (10), while the determination of the mass differences is not expected to improve significantly. This assumes $50$ fb$^{-1}$ LHCb and $50$ ab$^{-1}$ Belle II data, which may be achieved by the mid $2020$'s at the earliest \cite{Charles:2013aka}.

Implications of these measurements in terms of bounds for the specific $C+P$, $P$, $C$, and $CP$ symmetric LRSMs will be discussed in sections \ref{secCP1massmixing}, \ref{Pcase},\ref{Ccase} and \ref{CPcase}, respectively. 

\subsection{Electric dipole moments}
As mentioned before, LR models introduce a number of additional $CP$-violating sources. At low energies these will generally contribute to electric dipole moments (EDMs). In the lepton sector this leads to a nonzero electron EDM while $CP$-violating interactions in the quark sector EDMs can induce the EDMs of the neutron, proton and light nuclei. In the following we discuss the resulting bounds.
\subsubsection{Hadronic EDMs}
Hadronic EDMs receive contributions from the $CP$-violating phases in the CKM matrices and $\al$ as well as the QCD-theta term, $\bar \theta$. In a general LRM the latter is a free parameter. We will first discuss the case where $\bar\theta=0$, simply assuming this has been achieved through the implementation of a Peccei-Quinn mechanism or in some other way. Note, however, that this is not always the case, in fact, there is an interesting scenario in which $\bar \theta$ becomes calculable, leading to strong constraints on the LR scale \cite{Maiezza:2014ala}.

When $\bar\theta=0$ hadronic EDMs are dominated by a single interaction which appears at tree level while other contributions only appear at the loop level \cite{Xu2010,An:2009zh}. At the scale of $\sim 1$ GeV the operator responsible is given by
\bea\label{mLRSM1GeV}
\vL_{LR}&=&   -i \frac{2}{v\sq}\frac{g_R}{g_L}\sin\zeta\,\,\text{Im}\left(e^{i\al} V_L^{ud*}\,V_R^{ud} \right)\bigg[\eta_1 \left(\overline u_R \ga^\mu d_R\,\overline d_L \ga_\mu u_L -\overline d_R \ga^\mu u_R\,\overline u_L \ga_\mu d_L\right)\\
&&+ \eta_8 \, \left(\bar u_{R}\ga^\mu t_a d_{R}\, \bar d_{L}\ga_\mu t_a u_{L}-
\bar d_{R}\ga^\mu t_a u_{R}\, \bar u_{L}\ga_\mu t_a d_{L}\right)\bigg],\nn
\eea
where $t_a$ are the $SU(3)_c$ generators and $\eta_1 = 1.1$ and
$\eta_8 =1.4$ are QCD RGE factors \cite{Dekens:2013zca}. This operator is induced by tree-level exchange of $W_L^\pm-W_R^\pm$ bosons, see Fig.\ \ref{diagrams}c. Nonperturbative techniques are required to determine the contribution of this operator to the neutron EDM. Using naive dimensional analysis (NDA) \cite{NDA,Weinberg1989} and the upper limit on the neutron EDM, $d_n\leq 2.9\cdot 10^{-26}\, e\text{ cm}$ \cite{Baker:2006ts} one finds
\bea
\big|\frac{g_R}{g_L}\sin\zeta\,  \text{Im}\big(V_L^{ud*}V_R^{ud}e^{i\al}\big) \big|\leq 4\cdot 10^{-6},\label{nEDM}
\eea
with a considerable theoretical uncertainty. This is about a factor $40$ weaker than the upper bound found in Ref.~\cite{Zhang2008}, however, a recent analysis \cite{Seng14} using chiral perturbation theory ($\chi$PT) indicates that the constraint obtained there may have been overestimated. A stronger bound on the same combination may be derived from the limit on the mercury EDM \cite{Engel:2013lsa}, however, large nuclear uncertainties now play a role. In fact, taking the estimated uncertainties \cite{Engel:2013lsa} at face value, the contribution of the operator in Eq.\ \eqref{mLRSM1GeV} to the mercury EDM is consistent with zero.

It is also interesting to consider, in LR models, the EDMs of the proton, deuteron and helion ($^3$He), for which there are plans for measurements in storage rings \cite{Farley:2003wt, Onderwater:2011zz,Pretz,JEDI}. Using NDA estimates, one would expect the proton and neutron EDMs to be of similar size while the deuteron EDM is enhanced by about one order of magnitude \cite{Vries2013}. This implies the deuteron EDM is a more sensitive probe of LRMs than the neutron and proton EDMs.

Furthermore, although the lack of knowledge of the nonperturbative physics does not allow for a prediction of the absolute size of these EDMs, it is possible to relate them. This is due to the fact that the dominant contributions come from a single operator whose chiral symmetry properties imply these EDMs are not independent. An LRM in which the operator of Eq.~\eqref{mLRSM1GeV} is indeed dominant would predict \cite{Dekens:2014jka}\footnote{It should be noted however that there exists a caveat in the form of a three-pion interaction induced by Eq.~\eqref{mLRSM1GeV} which can also contribute to the tri-nucleon EDMs, possibly spoiling such a relation, see Ref.~\cite{Dekens:2014jka} for  details.},
\bea
d_{^3\text{He}} = (0.78\pm 0.18)\, d_D+(0.11\pm 0.24)\, d_n- (0.82\pm0.35)\,d_p,
\label{EDMrelation}\eea
where the errors are mainly due to nuclear uncertainties. Thus the measurements of the EDMs of light nuclei would provide a test of LR models. 

The above no longer applies when $\bar \theta\neq 0$. When $\bar\theta$ is a free parameter it becomes hard to say something in general about hadronic EDMs. A (partial) cancellation between the $\bar \theta$ contribution and that of the operator in Eq.~\eqref{mLRSM1GeV} can weaken or even evade the bound of Eq.~\eqref{nEDM}.
However, 
the $P$ and $CP$ symmetries in principle forbid $\theta$, in which case $\bar \theta$ is induced by the $CP$-violating phases in the quark mass matrices. 
$\bar \theta$ then becomes calculable in terms of $r\sin\al$ and Yukawa couplings \cite{Maiezza:2014ala}. For the $P$-symmetric case this means that instead of the operator in Eq.~\eqref{mLRSM1GeV} $\bar\theta$ gives the dominant contribution to the nEDM by far. The result is a very strong bound on $r\sin\al\lesssim \frac{m_b}{m_t}\cdot 10^{-10}$ which in turn implies (through $\varepsilon'$) a strong bound on $M_2\gtrsim 20\, \text{TeV}$ \cite{Maiezza:2014ala}. 

In some scenarios this bound on $\al$ also has consequences for the leptonic Yukawa couplings as they contribute to $\al$ at loop level. Assuming the Dirac Yukawa couplings for the quarks and leptons are similar the leptonic Dirac phase should naturally be $\lesssim 10^{-3}$, which would suppress $CP$ violation in neutrino oscillations beyond the reach of upcoming experiments \cite{Kuchimanchi:2014ota}. 

However, it is possible to suppress $\bar\theta$ and thereby the contribution to the nEDM, for example, by implementation of the Peccei-Quinn mechanism. The strong bounds on $\al$ and $M_2$ then no longer apply, the bounds that can be derived instead will be discussed in section \ref{Pcase}.

Returning to the $\bar\theta=0$ case, from the comparison of Eqs.\ \eqref{nEDM} and \eqref{phis} it is clear that the neutron EDM and the phases $\phi_{d,s}$ probe different combinations of $CP$-violating phases. The $B$-mixing observables, $\phi_{d,s}$, depend on the phases in $V_{L,R}$ while the neutron EDM is also sensitive to $\al$. Furthermore, the neutron EDM only receives contributions from $W_L$-$W_R$ mixing (Fig.\ \ref{diagrams}c) and thus depends on $\zeta$.  The best model independent bound on $\zeta$ allows a maximal value of $0.02$ \cite{Dunnen:2013wta}.\footnote{To be precise this bound holds for the combination $\text{Re}\big[\tan\zeta e^{i\al}\frac{g_RV^{ud}_R}{g_LV^{ud}_L}\big]\leq 0.02$. Limits of this order of magnitude were already derived in Ref.\ \cite{Bigi:1981rr} using hyperon decays, while more stringent limits $\Or(10^{-3})$ can be derived if one is willing to make certain assumptions about the CKM matrices \cite{Dunnen:2013wta}.} Barring cancellations between $\alpha$ and the phases in the right and left CKM matrices, $\zeta\sim 0.01$ will require $\sin\alpha < 10^{-4}$, but if $\zeta$ is smaller, $\sin \alpha$ is of course allowed to be larger. Instead, $\sin\phi_{d,s}$ receive contributions from box-diagrams and flavor-changing Higgs exchange (diagrams a and b of Fig.\ \ref{diagrams}), resulting in an $M_2$ and $M_H$ dependence. The $CP$ violation in Kaon mixing $\varepsilon$ is similar to $\phi_{d,s}$ concerning the dependence on the model parameters while $\varepsilon'$ is also sensitive to $\al$. This emphasizes the importance of the different precision measurements of $CP$ violation in order to probe all aspects of $CP$ violation in LR models.  

Thus, the flavor-diagonal $CP$ violation in the neutron EDM would seem to be complementary to the flavor-changing $CP$ violation appearing in meson mixing. Nonetheless, as we will discuss in upcoming sections, there are some scenarios in which these observables are no longer independent. Such correlations then lead to strong bounds on the right-handed scale.  
\subsubsection{The electron EDM}
Measurements of the electron EDM (eEDM) have recently improved considerably and also lead to a strong bound, at present $d_e\leq 8.7 \cdot 10^{-29} e\,\text{cm}$ \cite{Baron:2013eja}\footnote{What is probed in these experiments is actually a combination of the eEDM and semi-leptonic four-fermion interactions. The semi-leptonic interactions originate from tree-level diagrams which involve non-SM Higgs fields and thereby small Yukawa couplings. This, together with the fact that these Higgs fields should be heavy, of order $>10$ TeV for LRSMs, see section \ref{CP1Higgs}, implies that $d_e$ will generally  dominate.}. However, the eEDM is sensitive to other parameters than the hadronic EDMs, in this case the phases of the neutrino mixing matrix enter. Thus, the eEDM and hadronic EDMs are complementary observables. Furthermore, in principle the coupling of the right-handed bosons in the lepton sector may differ from those in the quark sector. Upon demanding anomaly cancellation one can relate the couplings from one sector to the other, but this could be altered by an as yet undiscovered fourth generation. 

Another difference is that for the eEDM there are no contributions from a leptonic equivalent of the four-quark operators in Eq.~\eqref{mLRSM1GeV}. This means that there are no tree-level contributions and the eEDM is generated at loop level. The generated eEDM is given by \cite{Nieves:1986uk, Chen:2006bv}
\bea
d_e\simeq -\frac{e^3}{16\pi^2 M_W\sq}\frac{g_R}{g_L}\sin\zeta \,\text{Im}\big(e^{-i\al} (M_{\nu_{D}})_{ee}\big),
\eea
where $(M_{\nu_{D}})_{ee}$ is the $ee$ element of the  neutrino Dirac-mass-matrix. It is in general not possible to compare the electron EDM to the neutron EDM as the two involve different phases. Nonetheless, we can still try to estimate their relative sizes. Taking the different phases to be of the same order and assuming $|(M_{\nu_{D}})_{ee}|\simeq m_e$ one finds $d_e/d_n\sim 10^{-4}$ (again assuming $\bar\theta = 0$) \cite{Dekens:2014jka}. 
  
\subsection{Neutron $\beta$ decay}
Bounds on LR models can also be obtained from neutron $\beta$ decay (n$\beta$d). In fact, Ref.\ \cite{Garcia:2010zza} even claims that it already provides evidence against the {\it manifest} LRSM \cite{Beg:1977ti}. Although not statistically significant, their result does show that especially the neutrino-neutron spin asymmetry $\alpha_\nu$ is very sensitive to the mass $M_2$. Here  $\alpha_\nu = 2[N(\theta_\nu <\pi/2) - N(\theta_\nu >\pi/2)]/[N(\theta_\nu <\pi/2) + N(\theta_\nu >\pi/2)]$, where $\theta_\nu$ is the angle between the neutrino direction and the polarization direction of the neutron. The analysis of \cite{Beg:1977ti} assumes that the right-handed current couples equally to leptons and quarks, which is an implicit assumption on the existence and mass of the right-handed neutrinos, namely that the decay to right-handed neutrinos is kinematically allowed. This requires that they should be light ($m_{\nu_R}\leq m_N$), which is not in accordance with a see-saw mechanism such as in the minimal LR models discussed here.

Neutron $\beta$ decay is sensitive to the CP-violating phase of the mixing between the $W_1$ and $W_2$ bosons in a similar but not completely identical way as the neutron EDM. Due to hadronic uncertainties which are hard to improve upon, the best current bound from n$\beta$d is not as strong as that of the neutron EDM. Assuming $\bar\theta=0$ and assuming heavy right-handed neutrinos, the best bound from n$\beta$d is\footnote{This bound is altered if we assume the right-handed neutrinos to be light. Again, this is not what one would expect in a LRM using the scalar triplets of Eq.\ \eqref{scalars}, however, this can be achieved in a LRM using doublets instead\cite{Senjanovic1979}.}
\begin{equation}
{\rm Im}\left( \tan \zeta e^{i\alpha} \frac{g_R}{g_L} \frac{V_R^{ud}}{V_L^{ud}}\right) = (1.0\pm 2.4)\cdot 10^{-4} \quad (68\%\,\text{CL}),
\end{equation}
which for LRSMs and small mixing angles translates into 
\begin{equation}
\frac{\kappa\kappa^\prime}{v_R^2} \sin(\alpha+\theta_u+\theta_d ) = (1.0\pm 2.4)\cdot 10^{-4}.
\end{equation}
This bound is obtained from Ref.\  \cite{Dunnen:2013wta} using updated experimental results \cite{Mumm:2011nd,Chupp:2012ta} and a lattice determination of $g_A/g_V=1.20(6)(4)$ \cite{Yamazaki:2008py}. For LRSMs the neutron EDM bounds the same quantity (also assuming $\bar\theta=0$) and is much stronger, see Eq.~\eqref{nEDM}, thus for LRSMs this bound is superseded by the nEDM constraint. However, for more general LRMs a comparison of the two observables would be sensitive to a deviation from $|V^{ud}_L|=|V^{ud}_R|$. Another way to detect deviations from the LR symmetric case is proposed in Ref.\ \cite{Fowlie:2014mza} which shows that a study using $b$-tags at the LHC would be sensitive to deviations of $|V_R^{tb}|$ from $|V_L^{tb}|$.
\newline

Having discussed the LR model and parts of its Lagrangian rather generally, we will now discuss in more detail the models which have an unbroken discrete symmetry, $P$, $C$, and/or $CP$ at high energies. Starting with the most symmetric option, we will discuss in the next section the LR models that  are both $P$ and $CP$ symmetric. 

\section{$P$- and $CP$-symmetric LR models}\label{C&Psym}
A LR model with both discrete symmetries has the appealing feature that $P$, $C$, and $CP$ violation are explained as low-energy phenomena.
However, there is no unique LR theory with both a $C$ and a $P$ symmetry as there are several ways of implementing both LR symmetries in Eq.~\eqref{C&Ptransf}. This is due to the fact that the $P$ and $C$ transformations need not be aligned in flavor-space \cite{Ecker:1983hz}. In what follows we will briefly discuss all possible ways of implementing both the $P$ and $C$ symmetries of Eq.~\eqref{C&Ptransf}, for a more detailed discussion see Refs.~\cite{Ecker:1983dj,Ecker:1985vv}.

Whenever the $C$ and $P$ symmetries are not aligned the transformation rules for $P$ or $C$ will not have the simple form of Eq.~\eqref{C&Ptransf}. 
We will select the basis in which the $P$ transformation does have the form of Eq.~\eqref{C&Ptransf}, while the $C$ transformation may be different. The implications in Eq.~\eqref{implications} for the $P$ symmetry are then unchanged so that $\Gamma$ and $\TG$ will be hermitian matrices. Following Ref.~\cite{Ecker:1983hz} we next demand invariance under $CP$ symmetry. 

\begin{figure}[t!]
\centering
\includegraphics[width=150mm]{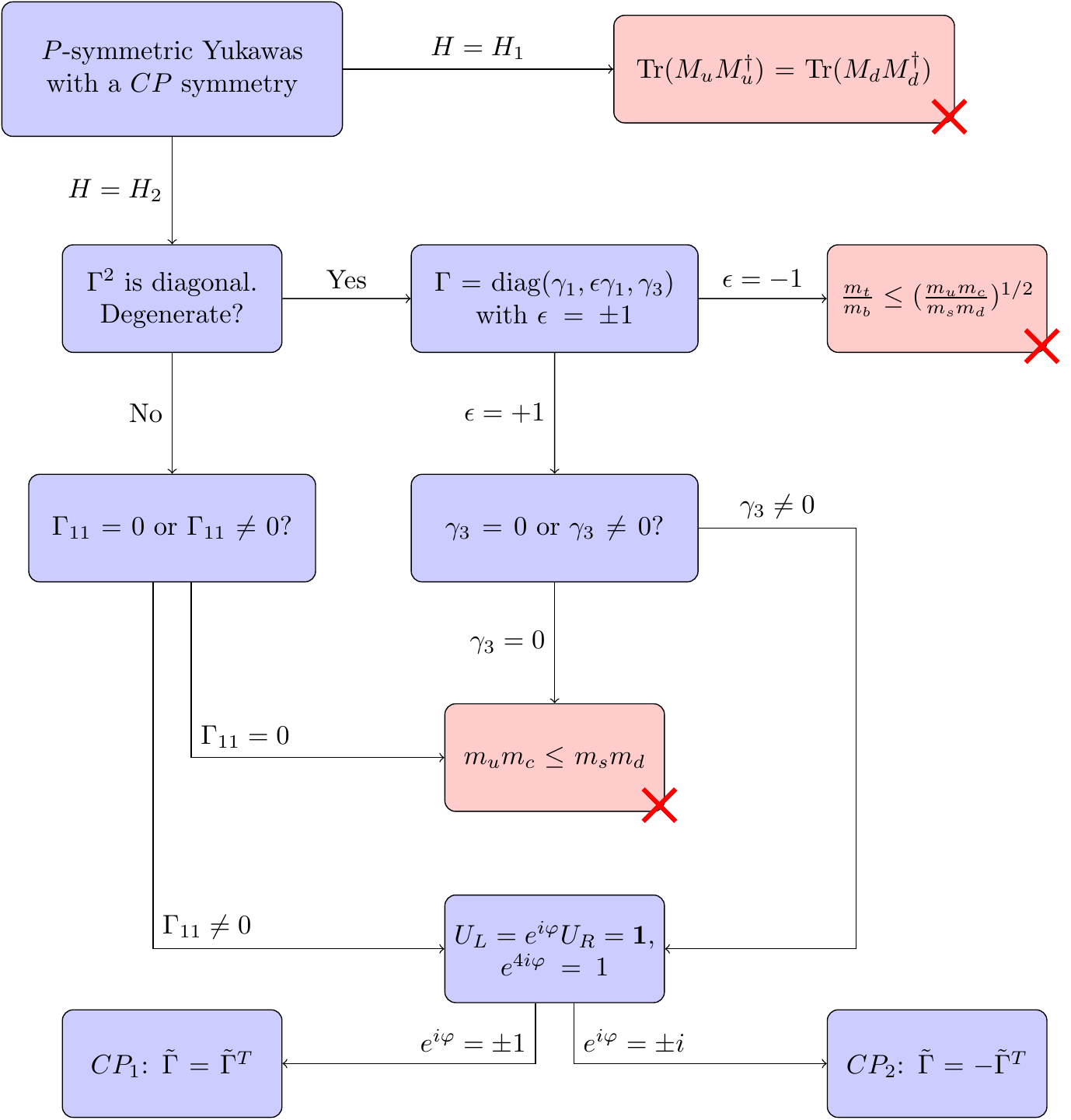} 
   \caption{Flowchart for a $P$- and $CP$-symmetric Yukawa sector, depicting the possible choices for the $CP$ transformation, Eq.~\eqref{CPtransf}, and their consequences. We work in a basis where $\Gamma$ is diagonal. See Refs.~\cite{Ecker:1983dj,Ecker:1985vv} for a detailed discussion.}
   \label{Flowchart} \end{figure}

The most general $CP$ transformation can be written as follows \cite{Ecker:1983hz}
\bea
Q_{L,R} \rightarrow U_{L,R} \,Q_{L,R}^{c}, \qquad \Phi \rightarrow H \Phi^* ,\qquad \Delta_{L,R} \rightarrow e^{i\phi_{L,R}}\Delta_{L,R}^*,
\label{CPtransf}\eea
where $U_{L,R}$ are unitary $3\times 3$ matrices, $\Phi \equiv (\phi,\, \tilde\phi)^T$, and $H$ is a unitary $2\times 2 $ matrix. As $\phi$ and $\tilde\phi$ are not independent fields this implies a relation between the elements of $H$. Taking this and the unitarity of $H$ into account there are two possible forms of $H$,
\bea
 H_1=\bma 0& \pm 1 \\\pm 1&0\ema,\qquad H_2=\bma e^{i\varphi} &0\\0&e^{-i\varphi}\ema ,
\label{Hforms}\eea
where $\varphi$ is a real number. These possibilities for $H$ and $U_{L,R}$ give rise to a number of possible $CP$ transformations, which in principle lead to different models. To simplify the discussion, and without loss of generality, we work in the basis where $\Gamma$ is diagonal. The possible transformation rules and the consequences of the resulting models are summarized in Fig.~\ref{Flowchart}. For a more detailed discussion see Refs.~\cite{Ecker:1983dj,Ecker:1985vv}. In short, only the option $H=H_2$ with $e^{4i\varphi}=1$ remains as a possible $CP$ transformation for the $\Phi$ fields, while the other possibilities are unable to reproduce the quark masses or their mixing. Thus, there are two $CP$ transformations which cannot be excluded on the basis of yielding unrealistic quark masses, namely \cite{Ecker:1983hz}
\bea
&CP_1: &\qquad H = \pm \mathbf{1}, \qquad U_{L} = \mathbf{1},\qquad U_{R} =\pm \mathbf{1},\nn\\
&CP_2: &\qquad H = \pm i\sigma_3, \qquad U_{L} = \mathbf{1}, \qquad U_R = \mp i \mathbf{1}.\label{CPtransf2}
\eea
We now discuss these two possibilities in more detail.

\subsection{$CP_1$-symmetric LR models}\label{secCP1}
\subsubsection{Mass and mixing matrices}\label{secCP1massmixing}
The $CP_1$ case is the more widely studied possibility, see for instance Refs.~\cite{Ecker:1983hz,Ecker:1985vv,Frere:1991db,Ball:1999mb}. At first sight, this model is able to produce phenomenologically viable mass and mixing angles for the quarks. However, as we will discuss, the model is unable to produce the observed $CP$ violation due to the $CP_1$ symmetry which constrains the Yukawa interactions as follows, 
\bea
\Gamma=\Gamma^T=\Gamma^*, \qquad \TG=\TG^T =\TG^*.
\eea 
The symmetric Yukawa couplings imply $V_R = K_u V_L^* K_d $ \cite{Branco:1982wp} for the CKM matrices, as in Eq.~\eqref{CKMC}. The hermiticity of the Yukawa couplings implies additional conditions which allows all phases in the CKM matrices to be solved in terms of known mixing angles, quark masses, and the parameters $\al$ and $r\equiv\kappa'/\kappa$ \cite{Chang:1982dp,Ecker:1985vv}. There are seven such phases to be solved which can be parametrized by the usual SM phase in the left-handed CKM matrix and six additional phases in the right-handed CKM matrix. 
This solution allows for the prediction of $CP$-violating observables in terms of the combination $r\sin\al$. However, in order to reproduce the quark masses this combination has to be small \cite{Maiezza:2010ic}, thus the above solution only exists for relatively small amounts of $CP$ violation, namely \cite{Frere:1991db,Ball:1999mb}
\bea 
\frac{ r\sin\alpha}{1-r\sq}\lesssim \frac{m_b}{m_t}. \label{CPsol}
\eea
The SM CKM phase $\dt$ can be expressed in terms of this combination as well, the above bound then requires this phase to be rather small, $\dt<0.25$ \cite{Ball:1999mb}. In contrast, the SM CKM fit requires $\dt$ to be rather large, $\dt\simeq 1.2$ \cite{pdg:2012}. This means that the $CP_1$ LR model cannot reproduce the $CP$ violation of the SM in the decoupling limit, $v_R\rightarrow \infty$. This limit can therefore be excluded \cite{Ball:1999mb}. 

The above observations have further implications. Since the decoupling limit does not reproduce the SM there is not necessarily a range in parameter space where the  $CP_1$ LR model reproduces the SM. In fact, from recent measurements of the $CP$ violation in $B-\bar B$ mixing the predictions of the model can be shown to be too small. After taking into account constraints from $CP$-violating parameters in the Kaon sector, $\varepsilon$ and $\varepsilon^{\prime}$, the model predicts (for any value of $M_2$ and $M_H$) \cite{Ball:1999yi,Frere:1991db},
\bea|\sin\phi_d|< 0.1,\qquad \sin\phi_s< -0.1. \label{phidPred} \eea
 Clearly, this is incompatible with the measured value in Eq.~\eqref{phid}. Thus, although the model can reproduce the observed quark masses and mixing angles, it is untenable when discussing $CP$-violating observables. As the Yukawa couplings of the pseudomanifest LRM coincide with that of the $P$- and $CP_1$-symmetric LRM under discussion here, it follows that the {\it minimal} pseudomanifest LRM can be excluded in the same way.
 
\subsubsection{The Higgs potential}\label{CP1Higgs}
Further problems arise when considering the Higgs sector of the $CP_1$ model. Although the above considerations may be considered to rule out the model we will nevertheless review some of these problems as they are exemplary of what will be encountered in other versions of LR symmetric models. Our discussion will largely follow that of Refs.~\cite{Barenboim:2001vu,Deshpande:1990ip}. 

In this case the Higgs potential takes the form of Eq.~\eqref{HiggspotP} with the additional constraint from the $CP_1$ transformation that there is no explicit $CP$ violation:
\bea \delta_{2} = 0.\eea
Note that if the phases $\phi_{L,R}$ are introduced in the $CP$ transformation of the $\Delta_{L,R}$ fields, Eq.~\eqref{CPtransf}, additional constraints, $\bt_i=\rho_4=0$, are acquired. However, as the potential Eq.~\eqref{HiggspotP} is $(B-L)$-symmetric no such constraints appear when $e^{i(\phi_{L}-\phi_R)}=1$. Such a Higgs potential has been widely studied in the literature \cite{Gunion:1989in,Basecq:1985sx,Deshpande:1990ip, Barenboim:2001vu,Kiers:2005gh}. From the requirement that the potential is minimized  one can obtain the following expressions for the dimensionful parameters of the potential \cite{Barenboim:2001vu}, 
\bea
\frac{\mu_1\sq}{v_R\sq} & =& \frac{\al_1}{2}-\frac{\ksq}{2\ka_-\sq}\al_3+\frac{\ka_+\sq}{v_R\sq}\la_1+2\frac{\ka\ka'}{v_R\sq}\la_4\cos\al, \nn\\
\frac{\mu_2\sq}{v_R\sq} &=& \frac{\al_2}{2}+\frac{\ka\ka'}{4\ka_-\sq \cos\al}\al_3 + \frac{\ka_+\sq}{2v_R\sq}\la_4+\frac{\ka\ka'}{v_R\sq\cos\al}(\la_3+2\la_2\cos2\al),\nn\\
\frac{\mu_3\sq}{v_R\sq} &=& \rho_1+\frac{\ka_+\sq}{2v_R\sq}\al_1+\frac{2\ka\ka'\cos\al}{v_R\sq}\al_2+\frac{\ksq}{2v_R\sq}\al_3,
\label{musol}\eea
where we neglected terms of order $v_L/ v_R$ and $v_L\sq/v_R\sq$. Substituting the exact expressions for $\mu\sq_i$ in the three remaining minimum equations we obtain
\bea
2\rho_1-\rho_3 &=& \frac{\bt_1\ka\ka'\cos(\al-\theta_L) +\bt_2\ka\sq\cos\theta_L+\bt_3\ksq\cos(2\al-\theta_L)}{v_R v_L},\label{minEq4}\\
0&=&\ka\ka'\big[\al_3(1+\frac{v_L\sq}{v_R\sq}) +\frac{\ka_-\sq}{v_R\sq}(4\lambda_3-8\lambda_2)\big]\sin\al \nn\\&& + \frac{ v_L}{v_R} \big[\bt_1\big(\ka\sq \sin\theta_L +\ka^{\prime 2}\sin(\theta_L-2\al)\big) + 2 (\bt_2+\bt_3)\kappa \kappa' \sin (\theta_L - \al)\big],\label{minEq5}\\
0&=&\bt_1\ka\ka'\sin(\al-\theta_L)-\bt_2\ka\sq\sin\theta_L+\bt_3\ksq\sin(2\al-
\theta_L).
\label{minEq6}
\eea
These equations were obtained by minimizing with respect to $v_L$, $\al$ and $\theta_L$, respectively. From Eq.~\eqref{minEq5} we approximately have (for $v_R\gg\ka_+\gg v_L$)
\bea\al_3\sin\al = \Or(\bt_i \,v_L/v_R).\eea
As we require the hierarchy $v_R\gg  v_L$, this implies that $\al_3\sin\al$ must be small in order for $\bt_1$ to be in the perturbative regime. 
As two extreme cases we could have a small spontaneous phase, i.e.\ $\al=\Or(v_L/v_R)$, while $\al_3$ can be of order one, or $\al_3$ is small with a sizeable $\al$, thus, $\al_3=\Or(v_L/v_R)$ with $\al=\Or(1)$. 
It turns out that in both extremes, as well as in all intermediate cases, some of the additional Higgs fields become too light, such that their effects should have been detected experimentally already \cite{Barenboim:2001vu,Deshpande:1990ip}. In one extreme, $\al=\Or(v_L/v_R)$ and $\al_3=\Or(1)$, 
Eqs.~\eqref{minEq4} and \eqref{minEq6} imply $2\rho_1-\rho_3 = \Or(\ka_+\sq/v_R\sq)$, which in turn implies that   
the left-handed triplet fields become light, $\Or(\ka_+)$. Explicit calculation shows that they are even lighter, namely of order $\Or(v_L)$. As these fields couple to the electroweak gauge-bosons such light fields should already have been discovered at LEP-I \cite{Barenboim:2001vu}. 
 
For the other extreme, whenever $\al_3$ is small, $\Or(\ka_+\sq/v_R\sq)$, there are problems with flavour-changing neutral-currents (FCNCs). In the minimal LR model these FCNCs are generated by the neutral scalars of the bidoublet. FCNCs are stringently constrained by Kaon- and $B$-mixing, in fact, the mass of such a scalar should be of the order of $>10$ TeV \cite{Ball:1999mb, Bertolini:2014sua} in LRSMs. An analysis of the masses of the physical Higgs fields \cite{Barenboim:2001vu} shows that whenever $\al_3$ is small, $\Or(\ka_+\sq/v_R\sq)$, the Higgs fields with the FCNC couplings remain light and the FCNC bounds cannot be evaded.

The remaining scenarios interpolate between the two extremes. Here one finds three light neutral states, which are now mixtures of the neutral triplet field, $\delta^0_L$, and the flavor-changing neutral Higgs field \cite{Barenboim:2001vu}. 

Thus, whenever $\sin\al \neq 0$ the potential Eq.~\eqref{HiggspotP} can not reproduce the SM Higgs spectrum. The addition of extra scalar fields could solve this, simultaneously allowing for a SM-like Higgs spectrum and spontaneous $CP$ violation in the $CP_1$ potential \cite{Branco:1985ng,Ball:1999mb}. This option will not be discussed any further here. Another possibility is to have a potential without spontaneous $CP$ violation, $\al = \theta_L = 0$. In this case all the non-SM scalars can obtain a large mass and decouple, thereby allowing for a SM-like Higgs spectrum. Even in this case, however, there is a price to pay in the form of fine-tuning as we will discuss next. 

\subsubsection{Fine-tuning}
The fact that there should be a hierarchy between the vevs, $v_R\gg\ka,\,\ka'\gg v_L$, induces fine-tuning in the potential. This occurs because the minimum equations relate the different scales to one another. For the parameters of the potential to be in the perturbative range this requires a certain amount of fine-tuning. Similar to the Higgs potential itself it is useful to review the fine-tuning in the $CP_1$ invariant potential, as it will turn out to be exemplary of the cases we will study in the following sections. 

One of the dominant sources of fine-tuning appears in Eq.~\eqref{minEq4}, schematically we have,
\bea
2\rho_1-\rho_3 \sim \frac{\ka\ka'}{v_L v_R}\bt_i,\label{vevseesaw}
\eea
which has been called the vev see-saw relation \cite{Deshpande:1990ip}. If we insist on the desired hierarchy, one may make the following estimates for the vevs, $v_R\sim 10$ TeV, $\ka, \ka'\sim 100$ GeV and $v_L\sim 1$ eV, which would imply the fraction in Eq.~\eqref{vevseesaw} to be huge $\sim 10^9$. Thus, in order for the $\rho$ parameters to be of order one, the $\bt_i$ parameters should cancel to a precision of  $\sim 10^{-9}$, implying a very fine-tuned potential. There are two ways to avoid this fine-tuning. One can either accept a very high scale for the LR model, $v_R\sim 10^{13} $ GeV, making the additional gauge fields, $Z_R$ and $W_R^\pm$ unobservable, or eliminate the vev see-saw relation. The latter option can be achieved by setting $v_L$ and $\bt_i$ to zero. In this case Eq.~\eqref{minEq4} vanishes and is no longer a source for fine-tuning. This option was concluded to be the only viable option leading to observable effects in Ref.~\cite{Deshpande:1990ip}. However, even in the case there is still a considerable amount of fine-tuning. Looking, for instance, at the third equation in Eq.~\eqref{musol} we see that the $\rho_1$ and $\mu_3\sq$ terms should cancel to $\Or(\frac{\ka_+\sq}{v_R\sq})$ in order for $\al_1$ to be of order one. Similarly, from the first equation in Eq.~\eqref{musol} the $\al_{1,3}$ and $\mu_1\sq$ terms should cancel to $\Or(\frac{\ka_+\sq}{v_R\sq})$ in order for $\la_i$ to be of order one. Combining the two, we see that cancellations to a precision of order $\Or(\frac{\ka_+^4}{v_R^4})$, i.e.\ $\Or(10^{-7}$) for the above selected values, are needed. Some of this fine-tuning may be avoided if we set some of the parameters to be small by hand or by introducing an additional mechanism \cite{Kiers:2005gh}.

As we will discuss later this type of fine-tuning tends to occur in more general scenarios as well. Before moving on to these LR models, however, we will first review the second $P$- and $CP$-symmetric case.

\subsection{$CP_2$-symmetric LR models}\label{secCP2}
This case has received somewhat less attention than the $CP_1$ possibility \cite{Ecker:1983hz,Branco:1985ng}. Although the Yukawa sector is distinct from the $CP_1$ case, which is relevant for the amount of $CP$ violation allowed, we will see that the Higgs potential is very similar.

\subsubsection{Mass and mixing matrices}
As in the previous case, the Yukawa interactions are constrained by the $P$ and $CP$ transformations. For the Yukawa interactions the $P$ and $CP_2$ transformations have the following implications
\bea
\Gamma = \Gamma^T= \Gamma^*,\qquad \TG=-\TG^T= -\TG^*.
\eea
The fact that $\TG$ is antisymmetric means that in this case the mass matrices will not be symmetric. Thus, there is, in general, no simple relation between the left and right-handed CKM matrices. However, the Higgs potential is clearly more constrained in this case. 

\subsubsection{The Higgs potential}
The $CP_2$ invariant Higgs potential is that of Eq.~\eqref{HiggspotP} with the additional constraints,
\bea 
&\mu_2=\lambda_4=0,\qquad \delta_{\al_2} =\pm\pi/2,&\nn\\
&\bt_1(1-e^{i(\phi_L-\phi_R)})=0,\qquad \bt_{2,3}(1+e^{i(\phi_L-\phi_R)})=0,\qquad \rho_4 (1-e^{2i(\phi_L-\phi_R)})=0.&
\eea
Thus, we have $\bt_1=0$ and/or $\bt_{2,3}=0$ (and possibly $\rho_4=0$) depending on $\phi_{L}-\phi_{R}$. 

As was already noted in Ref.~\cite{Branco:1985ng}, barring fine-tuning, there will be very little $CP$ violation in this case. Neglecting subleading terms in the potential, one finds $\al=\pm\pi/2$, which in this case is a $CP$ conserving minimum, as, after an $SU(2)_{L,R}$ gauge-transformation, the vevs of the bidoublet can then be written as 
\bea
\langle \phi\rangle  = e^{\pm i\pi/4}\bma \ka &0\\0&\ka'\ema,
\eea
which is invariant under the $CP_2$ transformation. 

The feature that without fine-tuning the spontaneous $CP$ violation will be small is reminiscent of the $CP_1$ case. In fact, we observe that the $CP_2$-symmetric potential is very similar to a special case of the $CP_1$-symmetric potential. This can be seen by use of a field redefinition. In case we take $e^{i(\phi_L-\phi_R)}=1$, and thereby $\bt_{2,3} = 0$, we can apply the following redefinition
\bea
\phi\rightarrow e^{\mp i\pi/4} \phi, \label{CP2redef}
\eea
the resulting potential is then, after an $SU(2)_{L,R}$ transformation, nearly equal to the $CP_1$ case with the following identifications,
\bea
&\mu^{CP_1}_2 = \lambda_4^{CP_1} =\bt_{2,3}^{CP_1}= 0,\qquad
\lambda_2^{CP_1} = -\lambda_2^{CP_2}, &\nn\\ &\al^{CP_1}=\al^{CP_2}\pm\pi/2 ,\qquad \theta_L^{CP_1}=\theta_L^{CP_2}\pm\pi/2.&
\label{CP2id}\eea
The only remaining difference comes from the $\al_2$ terms involving $\tr(\Delta_L\Delta_L^\dagger)$. Clearly, these terms are suppressed with respect to their right-handed equivalent, due to the hierarchy $v_R\gg \ka,\,\ka'\gg v_L$, meaning that to good approximation the two potentials are equivalent. Thus, in this case the minimum equations correspond to those of Eqs.~\eqref{musol} and \eqref{minEq4}-\eqref{minEq6} to $\Or(\ka_+\sq/v_R\sq)$,  with the identifications of Eq.~\eqref{CP2id}. A similar redefinition can be made for the case $e^{i(\phi_L-\phi_R)}=-1$.

Thus, the $CP_2$ potential is, to good approximation, equal to a special case of the $CP_1$ invariant potential. This also implies that the conclusions about the $CP_1$-symmetric potential carry over. The case with spontaneous $CP$ violation implies a non-SM-like Higgs spectrum whereas the case without spontaneous $CP$ violation has no $CP$ violation at all. Therefore, we conclude that also the $CP_2$-symmetric case is not viable.  
\newline

In the following sections we will study minimal LR models with fewer discrete symmetries. We will see that although there may be important differences between the Yukawa sectors of these models, their Higgs potentials will tend to be very similar, like for the $CP_1$ and $CP_2$ cases.

\section{$P$- or $C$-symmetric LR models}\label{CorPsym}

\subsection{$P$-symmetric LR models}\label{Pcase}
$P$-symmetric LRMs have been studied quite extensively in the literature, e.g. \cite{LRSM,Mohapatra:1974hk,Kiers:2005gh,Zhang2008,Bertolini:2014sua}. In this case there is an approximate relation between the mixing matrices, 
\bea V_L \simeq K_u V_R K_d, \label{CKMPapprox}\eea
where $K_{u,d}$ are diagonal matrices of phases. As was already mentioned, this is due to the fact that the combination $r\sin\al$ should be small in order to be able to reproduce the small ratio $m_b/m_t$ \cite{Maiezza:2010ic}. In fact, the same bound \eqref{CPsol} as in the $CP_1$ case applies here too. In order to satisfy this bound and simultaneously the experimental constraints on $CP$ violation, one can arrive at constraints on the phases in $K_{u,d}$ and on $M_2$. In other words, even though $CP$ violation can arise in this type of model, the pattern of $CP$ violation in Kaon and B-meson mixing and the nEDM may not be reproducible unless $M_2$ has some minimum value. Although this has been discussed before in the literature~\cite{Zhang2008, Bertolini:2014sua,Maiezza:2010ic}, we will briefly summarize this point. 

At the moment we do not know the value of $r$, but if it is small ($r\ll m_b/m_t$) one can use the analytical expressions for the phases in $K_{u,d}$ derived in terms of $r\sin\al$ \cite{Zhang2008}, which have recently been generalized to general $r$ values \cite{Senjanovic:2014pva}. For $M_2$ in the TeV range, the constraint on indirect $CP$-violation in Kaon mixing, $\epsilon$, then drives a combination of the phases in $K_{u,d}$ to a nonzero value, $|\theta_d-\theta_s| \simeq 0.17$. As in this case all the phases are functions of $r\sin\al$, this requires a nonzero value for this combination. Both the neutron EDM as well as $\varepsilon'$ then set a strong bound on $\zeta \sin\al$. These observables can then only be reconciled with $\varepsilon$ for large values of $M_2\gtrsim 10\,\text{TeV}$ \cite{Zhang2008,Bertolini:2014sua}.

On the other hand, if $r$ is large ($r\gtrsim m_b/m_t$) the phases in $K_{u,d}$ can be sizable and tuned to satisfy the $CP$ violation constraints. This leads to $\theta_c-\theta_t\simeq \pi/2$ from $\varepsilon$ and $\theta_d-\theta_s\simeq \pi$ from $\varepsilon'$. In addition, for the $CP$-violation in $B$ meson mixing (see Eq.~\eqref{phis}) we have
\bea
\sigma_d \simeq \pi+\theta_{b}-\theta_d, \qquad \sigma_s \simeq \pi+\theta_{b}-\theta_s,
\eea
which leads to a correlation between $\phi_d$ and $\phi_s$, on which we will comment further below \eqref{sigmads}. 
These constraints together with the Kaon and $B$-meson mass-differences, $\Delta M_K$ and $\Delta M_{B_{d,s}}$, require $M_2\gtrsim 3 \,\text{TeV}$ 
\cite{Bertolini:2014sua}. The recent analytical results of \cite{Senjanovic:2014pva} for general $r$ probably allows to put an even more stringent bound, 
as in this case all phases in $K_{u,d}$ are known expressions of $r\sin\al$, like in the $CP_1$ case\footnote{Note that in the $CP_1$ scenario also the SM phase $\dt$ can be expressed in terms of $r\sin\al$.}. 

In the limit of vanishing $\al$ the quark sector of the $P$-symmetric LRM coincides with that of the minimal manifest LRM. In this case $\varepsilon$ sets a strong bound on the $W_R^\pm$ mass of $M_2\gtrsim 20$ TeV \cite{Maiezza:2014ala}.

The above shows that the indirect constraints for $P$-symmetric LRMs are more stringent than the direct limits on $M_2$ and will be even more so in the future thanks to LHCb, $B$-factories and improvements in lattice determinations of the relevant matrix elements. The increase of experimental sensitivity discussed in section \ref{experiments}, which will take at least another $10$ years to realize, is expected to push the lower bound on $M_2$ to roughly $8$ TeV \cite{Bertolini:2014sua}, which will likely allow confrontation of the $P$-symmetric LRMs with data, although in this case there is no upper limit on $M_2$, as the decoupling limit has not been excluded, in contrast to the $CP_1$ case. 

\subsubsection{The Higgs potential}
The Higgs potential in this case is that of Eq.~\eqref{HiggspotP}. 
As pointed out in Ref.~\cite{Barenboim:2001vu} it can be mapped onto the $CP_1$ case to good approximation by a field redefinition, similar to that of Eq.~\eqref{CP2redef},
\bea
\phi \rightarrow \phi e^{-i\varphi/2}, \qquad \varphi  = \text{Arg} (\al_2 v_R\sq/2 e^{i\delta_2}  -\mu_2\sq ).
\label{PtoCP1}\eea
After an $SU(L)_L$-gauge transformation this gives
\bea
\al^{CP_1}\rightarrow \al^{P}+\varphi,\qquad \theta_L^{CP_1}\rightarrow \theta_L^{P}+\varphi  .\label{idenP}
\eea
The remaining differences between the two potentials then are terms subleading in $v_R$, $\Or(\ka_+\sq/v_R\sq)$. Thus, the minimum equations of Eqs.~\eqref{musol}--\eqref{minEq5} with the above replacement apply here too, to $\Or(\ka_+\sq/v_R\sq)$.
The remaining minimum equations result from subleading terms and are not obtained from their $CP_1$ equivalents after applying the identifications Eq.~\eqref{idenP}. Instead they simply equal the corresponding $CP_1$ equations,  Eqs.~\eqref{minEq4} and \eqref{minEq6}. This near-equivalence means the conclusions of Ref.~\cite{Barenboim:2001vu} about $CP_1$ models discussed in section \ref{secCP1} should apply here too.
Thus, again it will not be possible to obtain a SM-like Higgs spectrum for arbitrary values of $\al$. However, as mentioned in section \ref{secCP1} for the $CP_1$ case there is the possibility of a SM-like Higgs spectrum in the limit $v_R\rightarrow \infty$ for a specific value of $\al$ which now occurs very close to $\al = \varphi$. Note that this is now an acceptable possibility as it still allows for spontaneous $CP$ violation and we already allowed explicit $CP$ violation in the Yukawa couplings. Thus, in the $P$-symmetric LR model it is possible to have a SM-like spectrum in the decoupling limit in combination with spontaneous $CP$ violation. However, the amount of spontaneous $CP$ violation then entirely results from the explicit $CP$ violation present in the Higgs potential, as $\varphi = 0$ when $\delta_2 = 0$. In a sense this means that the $CP$ violation is put in by hand and can be as large as allowed by the value of $r$. Nevertheless, as discussed in the previous section, the pattern of $CP$ violation in Kaon and $B$-meson mixing and the nEDM also put stringent constraints on the model, in particular on $M_2$. Moreover, there is the issue of fine-tuning. 
 
\begin{table}[t]
\caption{\small The ranges taken for the parameters of the potential when generating random points in parameter space. }
\begin{center}\footnotesize
\begin{tabular}{||c|c|c|c|c|c|c|c||}
\hline
    \rule{0pt}{2.5ex}$\rho_1$&$\rho_3$& $\la_1,\,\rho_2,\,\al_{2,3}$ & $\la_{2,3,4},\,\al_1,\,\rho_4,\,\bt_{1,2,3}$& $\al,\,\delta_2,\,\theta_L$& $v_L$ (eV)&$\ka$ (GeV)& $v_R$ (GeV)\\
    \hline
$[0,5]$ &$[2\rho_1, 10]$&$[0,10]$ &$[-10,10]$ &$[0,2\pi]$ &$[0,10]$ &$[0,246 ]$ & $[0,5\cdot 10^4]$\\\hline
\end{tabular}
\end{center}
\label{FTranges} 
\end{table}

\begin{figure}[t!]
\centering
\includegraphics[width=75mm]{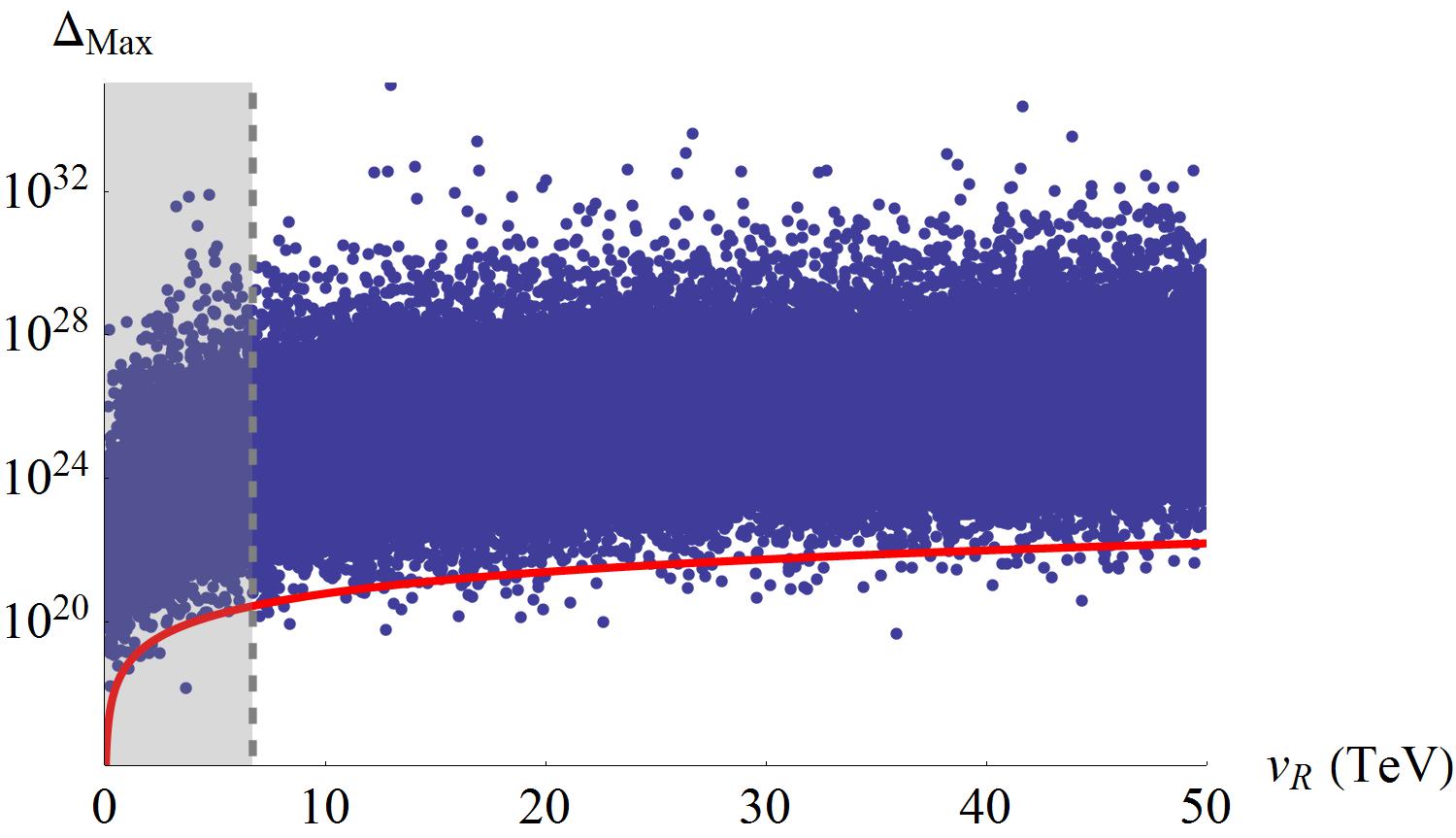} 
\caption{The figure shows the fine-tuning measure $\Delta_{\text{Max}}$ as a function of $v_R$ in TeV for a $P$-symmetric $V_H$. The blue points are randomly generated points satisfying the minimum equations and the ranges in Table \ref{FTranges}. Here the red line is chosen such that $0.1\%$ of the points are found below it. It is parametrized by
$6 \cdot 10^{-4} v_R^2/v_L\sq$ taking an average value for $v_L$ of $10$ eV.} 
\label{Pfinetune}
\end{figure} 

\subsubsection{Fine-tuning}
The fact that the minimum equations relate several very different scales, namely, $v_R\gg \ka_+\gg v_L$ to one another means that some of the parameters in the potential will tend to be fine-tuned. 
This is especially true for cases where $v_L\neq 0$ \cite{Deshpande:1990ip} as was already noted in section \ref{secCP1}. In this case one can obtain the vev see-saw relation of Eq.~\eqref{vevseesaw}. This requires some parameters to be fine-tuned to a precision of order $\Or(v_L/v_R)$. However, as we will show, more fine-tuning may be required, similar to that discussed in section \ref{secCP1}, due to the remaining minimum equations. In fact, solving the minimum equations  for $\mu_{1,2}\sq$, $\bt_{1,2}$, $\al_2$ and $\rho_3$ we see that the leading terms in $\rho_3 $ are proportional to $v_R\sq/v_L\sq$. This implies that if $\rho_3 $ is to be of order one, i.e.\ in the perturbative regime, these terms should cancel to a precision of $v_L\sq/v_R\sq$.

In order to study the matter quantitatively we use the minimum equations to solve for as many parameters, which we will denote by $p_i$, as there are equations. Subsequently we study the dependence of these $p_i$ on the remaining parameters, $p_j$. More specifically, we adopt the following quantity as a measure for the fine-tuning in $p_i$ typically used for supersymmetric extensions of the SM \cite{Ellis:1986yg,Barbieri:1987fn},
\bea
\Delta_i = \text{Max}_j\, \bigg|\frac{d \ln p_i}{d\ln p_j}\bigg|.
\label{FTdef}\eea
\begin{figure}[t]
\centering
$v_L=\bt_i=0$ \hspace*{55mm} $v_L=0,\,\bt_i\neq 0$
\\
\includegraphics[width=75mm]{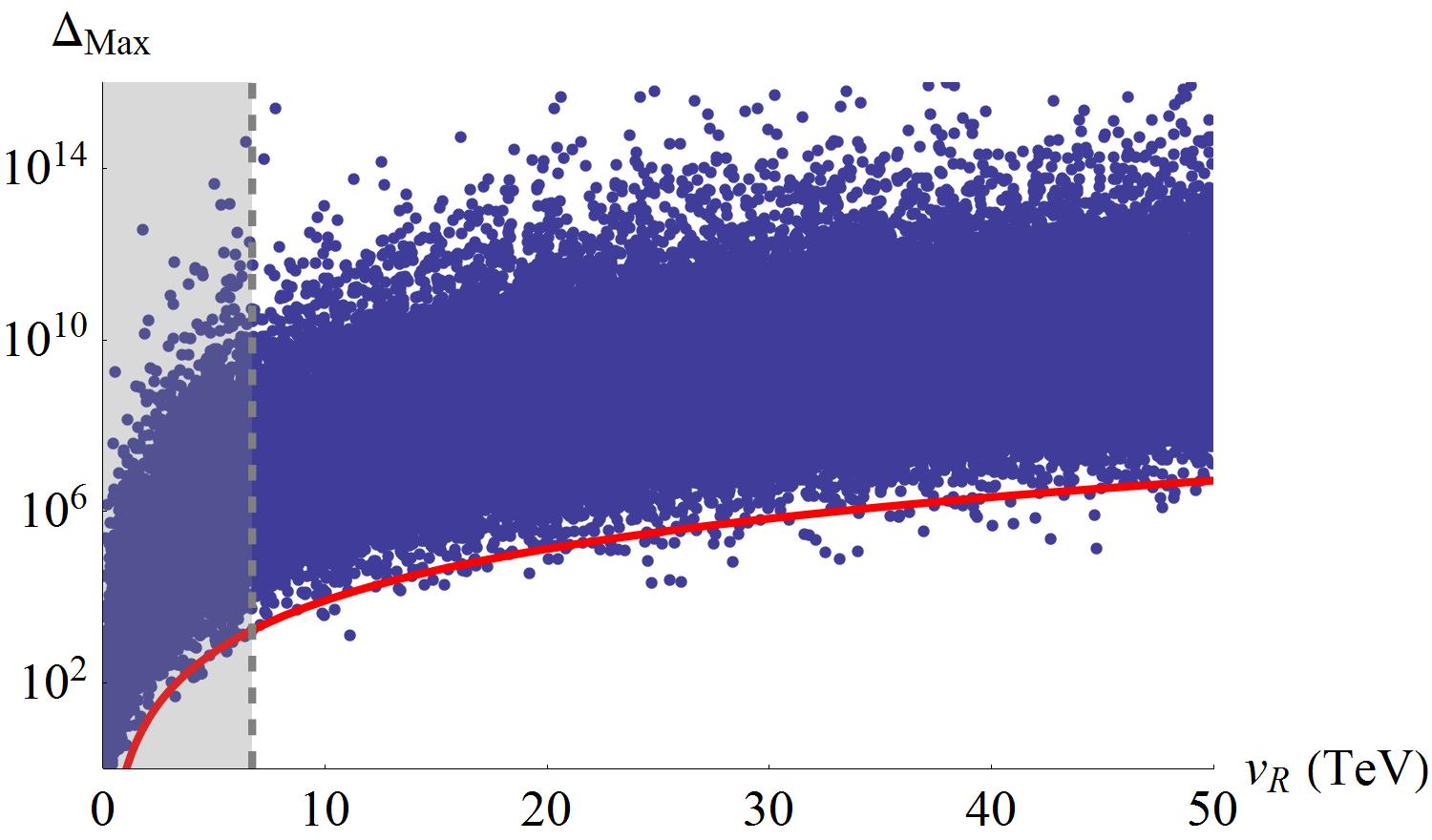} 
\includegraphics[width=75mm]{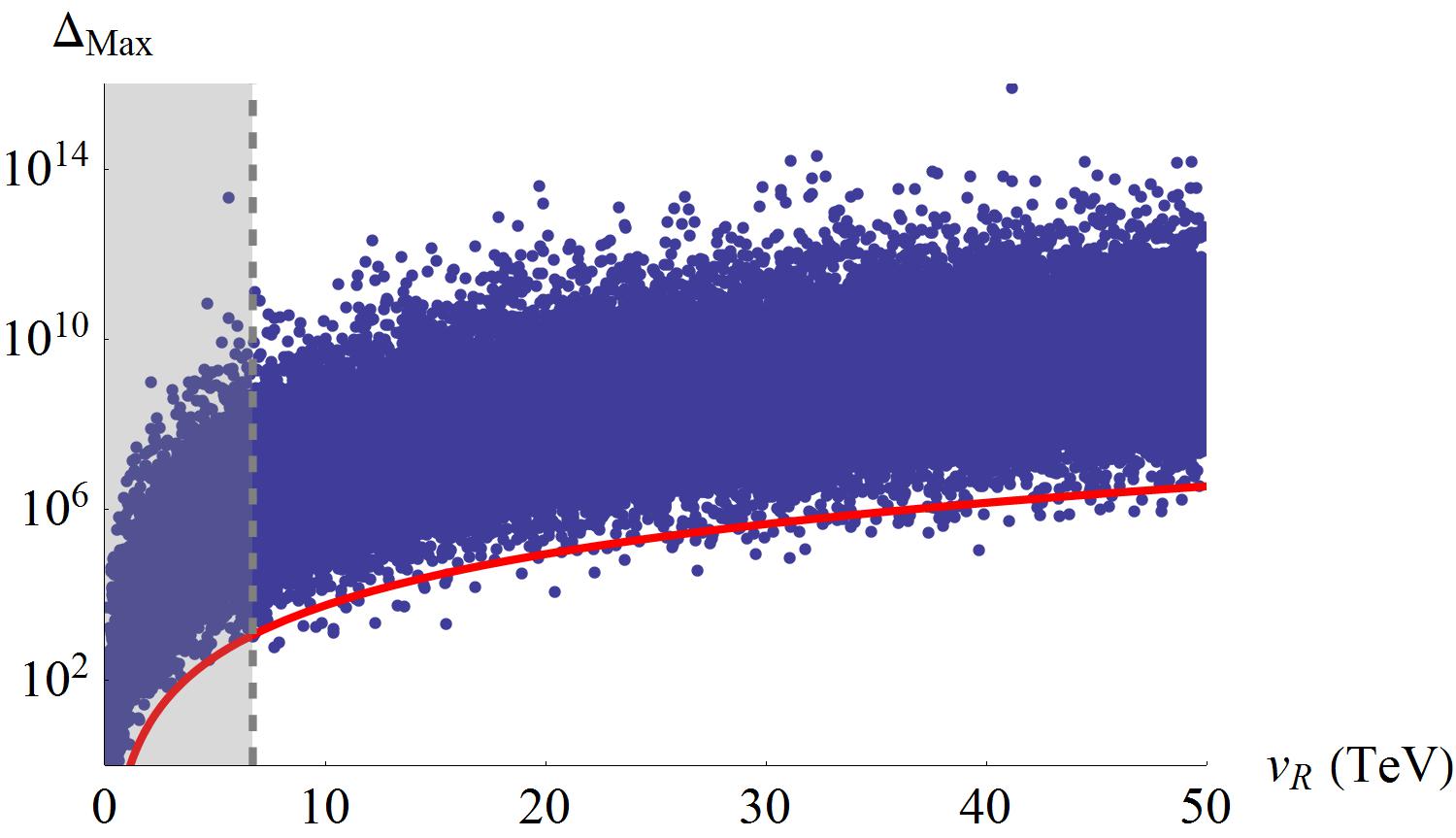} 
\caption{Similar plots to that of Fig.~\ref{Pfinetune}. The plot on the left shows the fine-tuning in the case where $\bt_i$ and $v_L$ are set to zero, while the figure on the right does the same in the case where only  $v_L$ is set to zero. The red lines are again chosen such that $0.1\%$ of the points are found below it. It is parametrized by $3 \cdot 10^{-3} v_R^4/\ka_+^4$
and $2 \cdot 10^{-3} v_R^4/\ka_+^4$ in the left and right plots, respectively.} 
\label{PfinetunevL=0}
\end{figure} 

Our procedure is as follows, we first generate random $\Or(1)$ values for nearly all parameters in the potential while obtaining values for the remainder through the minimum equations. The allowed ranges for the parameters are shown in Table \ref{FTranges}. As the dimensionless parameters should remain in the perturbative range we
conservatively constrain their values to lie in the interval $[-10,10]$. Exceptions occur for several dimensionless parameters. These are further constrained in order for the potential to be bounded from below, in the case of $\la_1$ and $\rho_1$ and to have positive masses-squared values of the Higgs fields, like for $\rho_2$, $\rho_3-2\rho_1$, and $\al_3$ \cite{Kiers:2005gh}. However, the imposed constraints are actually not sufficient to keep all mass-squared values positive, but this will not affect the conclusions. In addition, we have not imposed any experimental constraints on these masses. Thus, only a subset of the generated points will be phenomenologically viable. We further take values for the vevs which adhere to their naive expectations. We make no assumptions for the $\mu\sq_i$ parameters, instead we calculate their values through the minimum equations.

We then solve the minimum equations for as many parameters as there are equations. The random points in parameter space are then used to calculate the fine-tuning measures for the solved parameters $p_i$. The results are shown in Fig.~\ref{Pfinetune} where we plot the maximum value of $\Delta_\text{Max}\equiv \text{Max}_i \,\Delta_i$ against $v_R$.

Clearly, the degree of fine-tuning can be significantly larger than one might expect from the see-saw relation of Eq.~\eqref{vevseesaw} alone and be enhanced through the coupled minimum equations. As can be seen from the plot in Fig.~\ref{PfinetunevL=0} and was noted in Ref.~\cite{Deshpande:1990ip} the fine-tuning may be considerably decreased by setting $v_L=\bt_i=0$, as was done in e.g. \cite{Zhang2008}. In this case, however, still a fine-tuning of order $\Delta = \Or(v_R^4/\ka_+^4)\gtrsim 100$ remains. Since setting $\bt_i=0$ may lack justification \cite{Deshpande:1990ip,Barenboim:2001vu}, we observe that setting only $v_L$ to zero leads to the same reduction in the amount of fine-tuning. In this case the vev see-saw relation vanishes and instead we obtain two relations for the $\bt_i$ parameters, namely,
\bea
\bt_1 = -2\bt_3\frac{\ka'}{\ka}\cos\al,\qquad \bt_2 =\frac{\ksq}{\ka\sq} \bt_3. \label{betaRel}
\eea
It remains to be seen whether these relations can be justified or not. Therefore, setting $v_L$ and possibly $\bt_i$ to zero is a simple way to greatly reduce the fine-tuning in the Higgs potential, but in both cases one may wonder whether this is justified.

\subsection{$C$-symmetric LR models}\label{Ccase}
$C$-symmetric LRMs have been less investigated in the literature so far, although recently there has been renewed interest in Refs.~\cite{Maiezza:2010ic} and \cite{Bertolini:2014sua}. In this case the mixing matrices are related by Eq.~\eqref{CKMC},
\bea V_R = K_u V_L^* K_d,\eea
with $K_{u,d}$ diagonal matrices of phases. The mass matrices are now less constrained than in the $P$-symmetric case; the Yukawa couplings are symmetric as opposed to being hermitian. One of the consequences is that the combination $r\sin\al$ is no longer required to be small in order to reproduce  the quark masses, as opposed to the $P$-symmetric case. Furthermore, now the phases in $K_{u,d}$ are free parameters of the model. These can be tuned in order to evade the constraints from $CP$-violating observables. The $\varepsilon$ constraint can be evaded when $|\theta_d-\theta_s|\simeq n\pi$ ($n=0,1$), while the $\varepsilon'$ constraint can be satisfied when $\theta_d-\theta_s\simeq\pi$ or $r$ is small. From the relation in Eq.~\eqref{CKMC} we have for the phases $\sigma_{d,s}$ (see Eq.~\eqref{phis})
\bea
\sigma_d =\pi+ \theta_b-\theta_d+2\phi,\qquad \sigma_s =  \pi+\theta_b-\theta_s,
\label{sigmads}
\eea
where $\phi=\text{Arg}\big(V_L^{tb*}V_L^{td}\big)$. Note that $\sigma_{d}$ and $\sigma_{s}$ are again correlated in an $M_2$ and $M_H$ dependent way through the $\varepsilon^{(\prime)}$ constraints. To be specific, for the quantities which determine the $B_{d,s}$ observables one has
\bea
1+h_d \simeq 1-|h_d|e^{i(\theta_b-\theta_d+2\phi)},\qquad 1+ h_s \simeq 1-(-1)^n|h_s|e^{i(\theta_b-\theta_d)},\label{CBmixing}
\eea
while in the $P$-symmetric case, again taking into account $\varepsilon^{(\prime)}$ constraints, one has
\bea
1+h_d \simeq 1-|h_d|e^{i(\theta_b-\theta_d)},\qquad 1+h_s \simeq 1+|h_s|e^{i(\theta_b-\theta_d)}.\label{PBmixing}
\eea
From the $B_{d,s}$-mixing observables ($\Delta M_{B_{d,s}}$ and $\phi_{d,s}$) it should in principle be possible to see which of the above patterns, Eq.~\eqref{CBmixing} or \eqref{PBmixing}, fits the data better (especially since $|h_d|/|h_s|$ is constant to good approximation \cite{Ball:1999mb,Maiezza:2010ic}). In other words, if a sign of an LRSM is found in $B_{d,s}$-mixing these observables are in principle also sensitive to the difference between the $P$- and $C$-symmetric options. 

After the $\varepsilon^{(\prime)}$ constraints  have been used to fix the phases, the constraints from $\Delta M_K$ and from $B_{d,s}$ mixing can again be used to put a strong limit on the $W_R^\pm$ mass. In this case the $B_{d,s}$ meson limits are competitive with the bound from Kaon mixing \cite{Bertolini:2014sua} which is similar to the $P$-symmetric case, i.e.\ $M_2\gtrsim 3 \,\text{TeV}$.

\subsubsection{The Higgs potential}
\begin{figure}[t!]
\centering
\includegraphics[width=75mm]{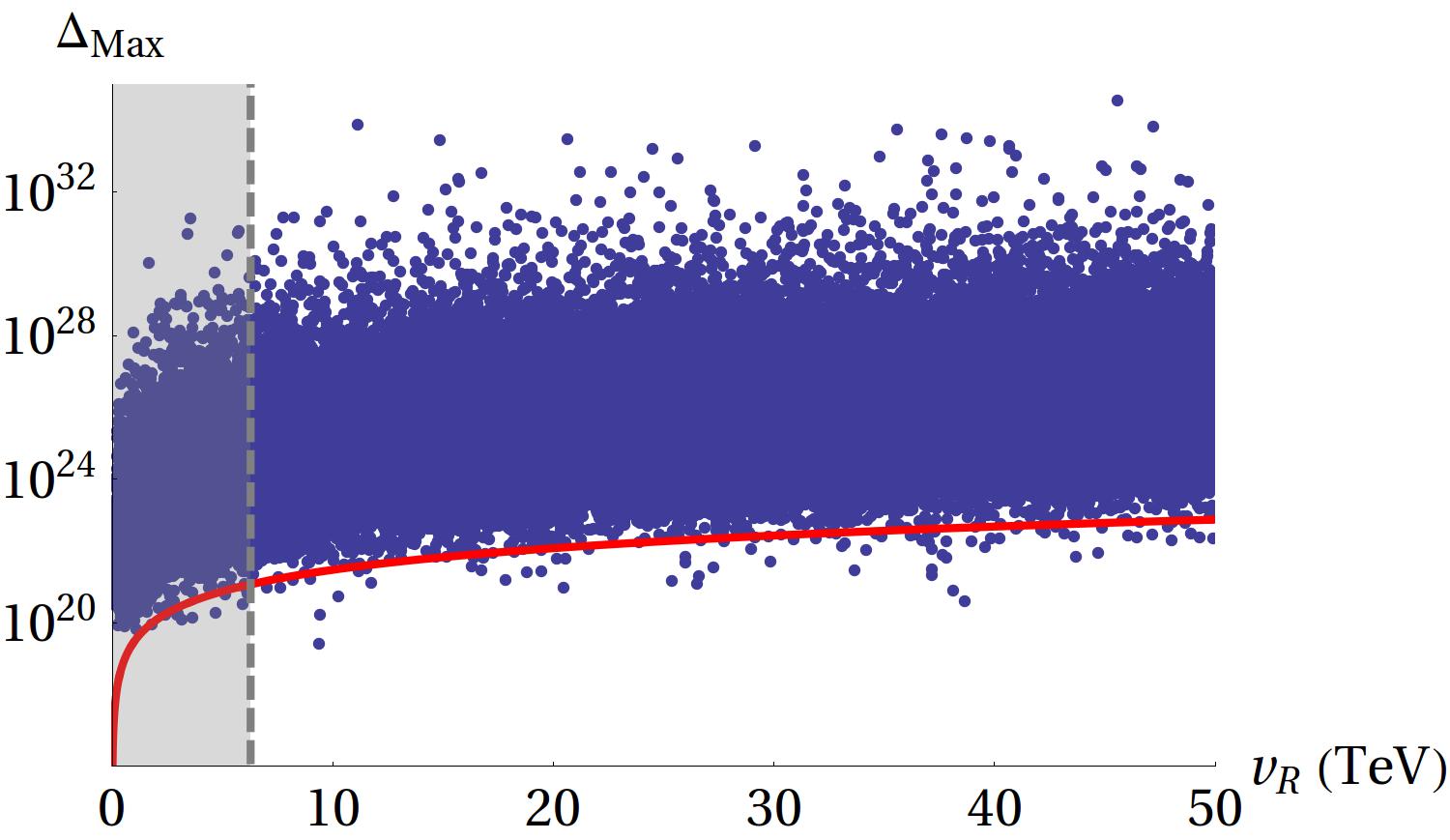} 
\caption{The figure shows the fine-tuning measure $\Delta_{\text{Max}}$ as a function of $v_R$ in TeV for a $C$-symmetric $V_H$. With respect to the $P$-symmetric potential the $C$-symmetric potential has seven additional phases, for which we choose the range $[0,2\pi]$. The red line is parametrized by $30 \cdot 10^{-4} v_R^2/v_L\sq$ taking an average value for $v_L$ of $10$ eV.} 
\label{Cfinetune}
\end{figure} 
The Higgs potential in this case is that of Eq.~\eqref{HiggspotC}. The form of this potential is quite  similar to that of the $P$-symmetric potential, Eq.~\eqref{HiggspotP}. In this case, however, phases  appear in the $\mu_2$, $\lambda_{2,4}$, $\rho_4$ and $\bt_i$ terms which were absent in the $P$-symmetric potential. Nonetheless, to good approximation the potential of Eq.~\eqref{HiggspotC} can again be mapped onto the $CP_1$-symmetric case by a field redefinition,
\bea
\phi \rightarrow e^{-i\varphi'/2 }\phi,\qquad \varphi'=\text{Arg}\big(\frac{\al_2v_R\sq}{2}e^{i\dt_{\al_2}}-\mu_2\sq e^{-i\dt_{\mu_2}}\big) .\label{Credef}
\eea
After an $SU(2)_L$-gauge transformation this then implies the following identifications,
\bea
 \al^{CP_1}=\al^{C}+\varphi ,\qquad \theta_L^{CP_1}=\theta_L^{C}+\varphi .
\label{CtoCP1}
\eea
This redefinition removes the phases from the $\mu_2\sq$ and $\al_2$ terms, but does not remove the phases related to the $\lambda_{2,4}$, $\rho_4$, and $\bt_i$ terms. However, these terms are subleading, i.e.\ of order $\Or(\ka_+\sq/v_R\sq)$ and smaller. At first sight the $\rho_4$ term may be an exception to this, however this term only contributes an $\Or(v_L v_R)$ term to the masses of doubly charged scalars and does not appear in the minimum equations. Thus, up to terms subleading in $v_R$, after the redefinition Eq.~\eqref{Credef} the $C$-symmetric potential is equal to the $CP_1$-symmetric potential. Indeed four of the minimum equations are again those of the $CP_1$ case, Eq.~\eqref{musol} and \eqref{minEq5}, to $\Or(\ka_+\sq/v_R\sq)$, with the above identifications. The remaining two, do not follow this rule as they emerge from subleading terms in the potential. Instead they are given by,
\bea
2\rho_1 -\rho_3 &=& \frac{\bt_1\ka\ka'\cos(\dt_{\bt_1}+\al-\theta_L)+\bt_2\ka\sq\cos(\dt_{\bt_2}-\theta_L)
+\bt_3\ksq\cos(\dt_{\bt_3}+2\al-\theta_L)}{v_Rv_L},\nn\\
0&=&\bt_1\ka\ka'\sin(\dt_{\bt_1}+\al-\theta_L)+\bt_2\ka\sq\sin(\dt_{\bt_2}-\theta_L)
+\bt_3\ksq\sin(\dt_{\bt_3}+2\al-\theta_L)
\eea
Nevertheless, up to small corrections the conclusions of $P$- and $CP_1$-symmetric case, discussed in section \ref{secCP1}, should apply once more.

The conclusions for the $C$-symmetric potential are very much like those for the $P$-symmetric case. Again it will not be possible to obtain a SM-like Higgs spectrum for arbitrary values of $\al$. Nonetheless, there is a possibility of a SM-like Higgs spectrum in the decoupling limit for a specific value of the spontaneous phase which in this case occurs for $\al = \varphi'$. Thus, much like the $P$-symmetric potential, it is possible to have a SM-like spectrum in the decoupling limit in combination with spontaneous $CP$ violation, however, the size of the spontaneous phase is again entirely dictated by the explicit $CP$ violation present in the potential, through Eq.\ \eqref{Credef}.
\subsubsection{Fine-tuning}
\begin{figure}[t]
\centering
$v_L=\bt_i=0$ \hspace*{55mm} $v_L=0,\,\bt_i\neq 0$
\\
\includegraphics[width=75mm]{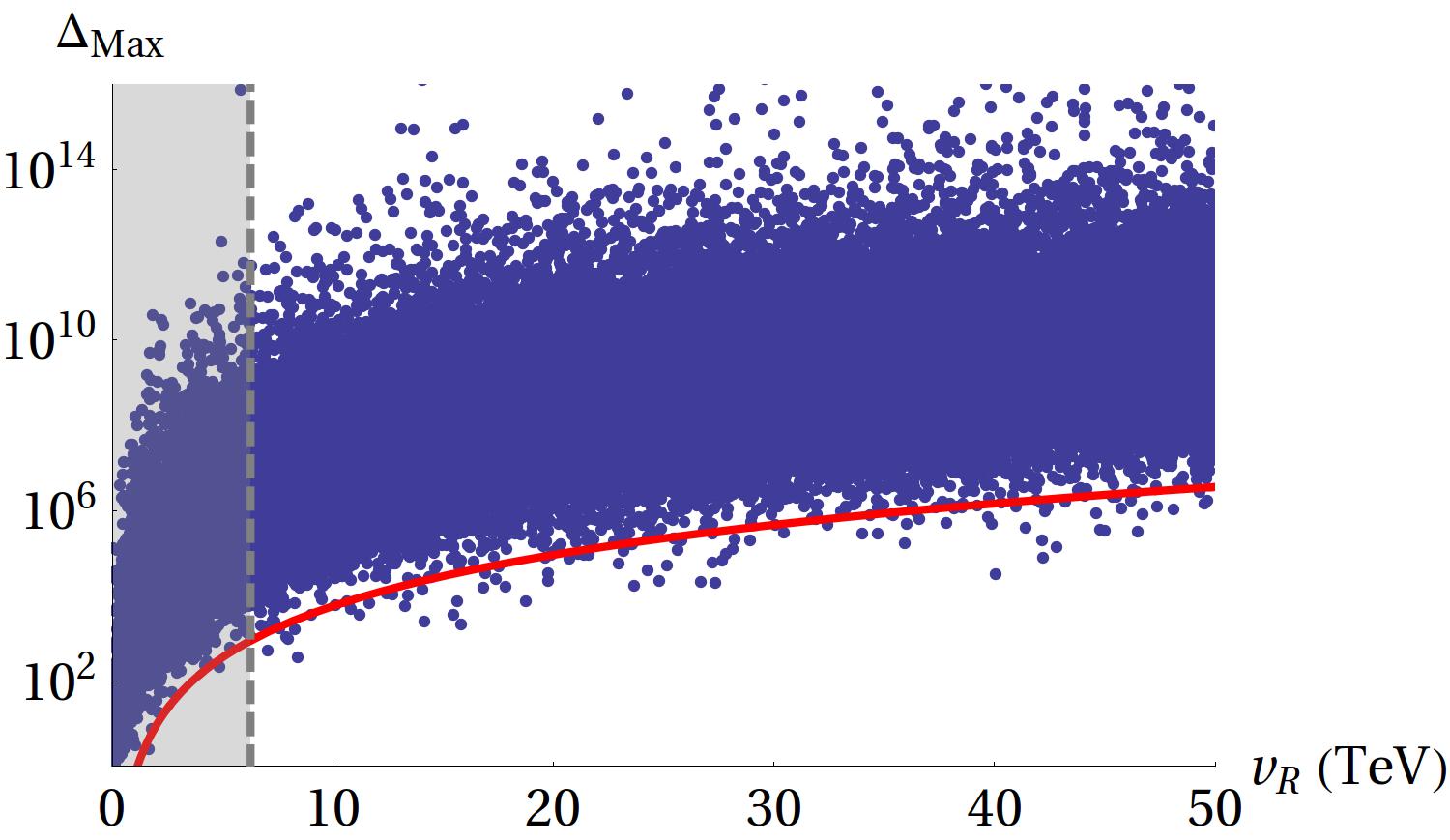} 
\includegraphics[width=75mm]{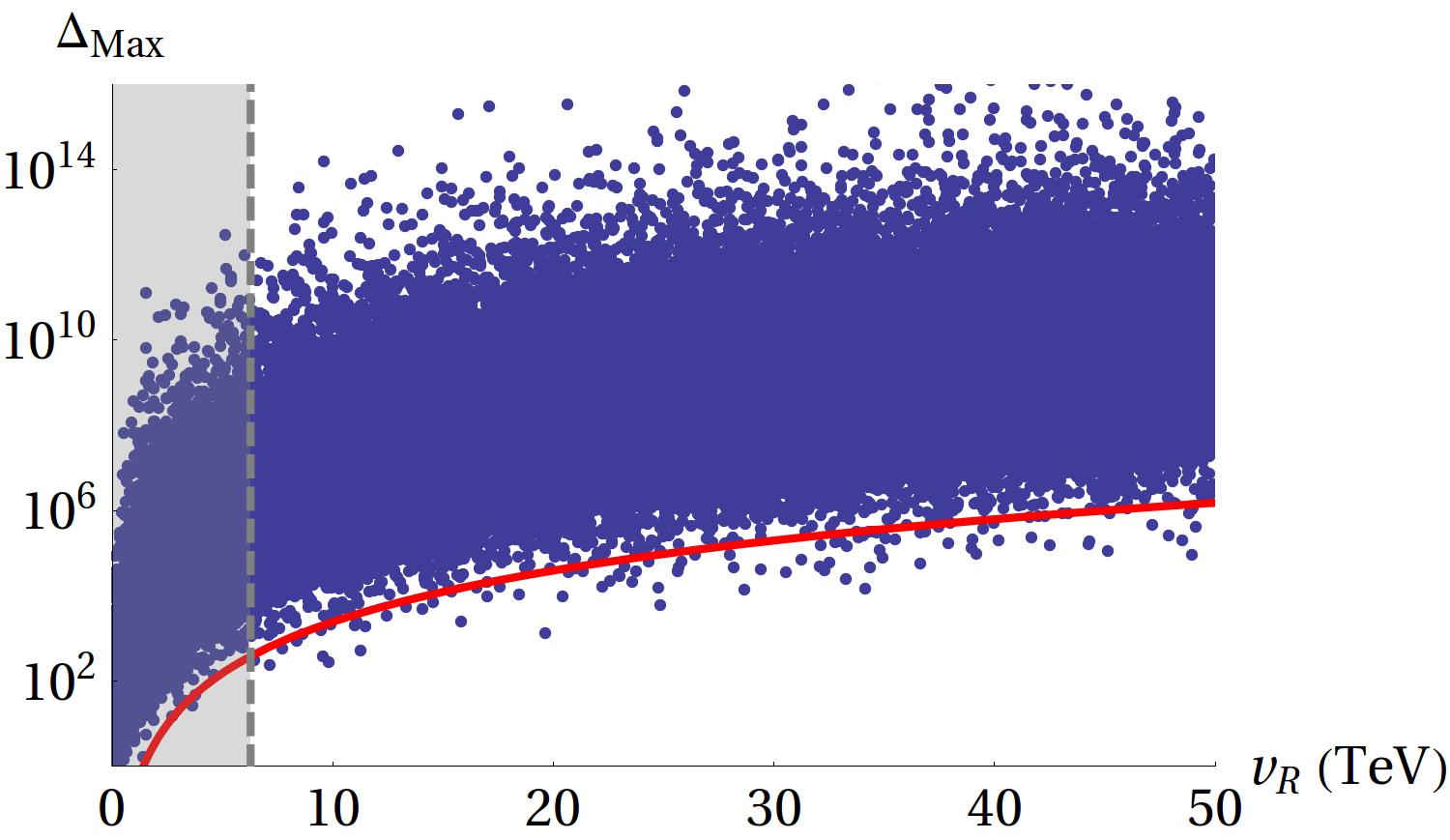} 
\caption{Similar plots to that of Fig.~\ref{Cfinetune}. The plot on the left shows the fine-tuning in the case where $\bt_i$ and $v_L$ are set to zero, while the figure on the right does the same in the case where only $v_L$ is set to zero. The red lines are again chosen such that $0.1\%$ of the points are found below it. It is parametrized by $2 \cdot 10^{-3} v_R^4/\ka_+^4$
and $1 \cdot 10^{-3} v_R^4/\ka_+^4$ in the left and right plots, respectively.} 
\label{CfinetunevL=0}
\end{figure} 
Not surprisingly, the fine-tuning measures in the $C$- and $P$-symmetric potentials are rather similar. If we do not eliminate the vev see-saw relation, Eq.~\eqref{vevseesaw},
the potential must again be very fine-tuned, as can be seen in Fig.~\ref{Cfinetune}, which can be compared with Fig.~\ref{Pfinetune} of the $P$-symmetric case. 

One can again choose to eliminate the see-saw relation in order to reduce the amount of fine-tuning substantially. As before, this can be achieved by setting $v_L=0$ and possibly $\bt_i=0$, see Fig.~\ref{CfinetunevL=0}. In both cases still a considerable amount of fine-tuning remains, $\Delta = \Or(v_R^4/\ka_+^4)\gtrsim 100$. In case of $v_L=0$ only, one obtains again two relations for the $\bt_i$ parameters, which however differ from those of Eq.~\eqref{betaRel},
\bea
\bt_1 = r\bt_3\frac{\sin(\dt_{\bt_2}-\dt_{\bt_3}-2\al)}{\sin(\dt_{\bt_1}-\dt_{\bt_2}+\al)}, \qquad \bt_2 = -r\bt_1.
\eea

\subsection{$CP$-symmetric LR models}\label{CPcase}
In a $CP$-symmetric LR model the CKM matrices are in principle unrelated to one another, the same is true for the gauge couplings, generally, $g_R\neq g_L$. Nevertheless, in a similar fashion to the $P$ and $CP_1$ cases \cite{Maiezza:2010ic} one may again derive the upper bound of Eq.\ \eqref{CPsol}. Despite this similarity, here it is possible to tune the right-handed gauge coupling and CKM elements in order to weaken the constraints from direct searches and $B$ and $K$ mixing.
In fact, the study of general LRMs (without the constraints on $V_R$ of the $C$ and $P$ cases) of Ref.\ \cite{Blanke:2011ry} shows that for $M_2$ in the $2-3$ TeV range a nearly diagonal form of $V_R$, much like that of the SM CKM matrix, is allowed, but also regions with large off-diagonal ($V^{cb}_R$ and $V_R^{ub}$) elements are possible.

\subsubsection{The Higgs potential}
\begin{figure}[t!]
\centering
\includegraphics[width=75mm]{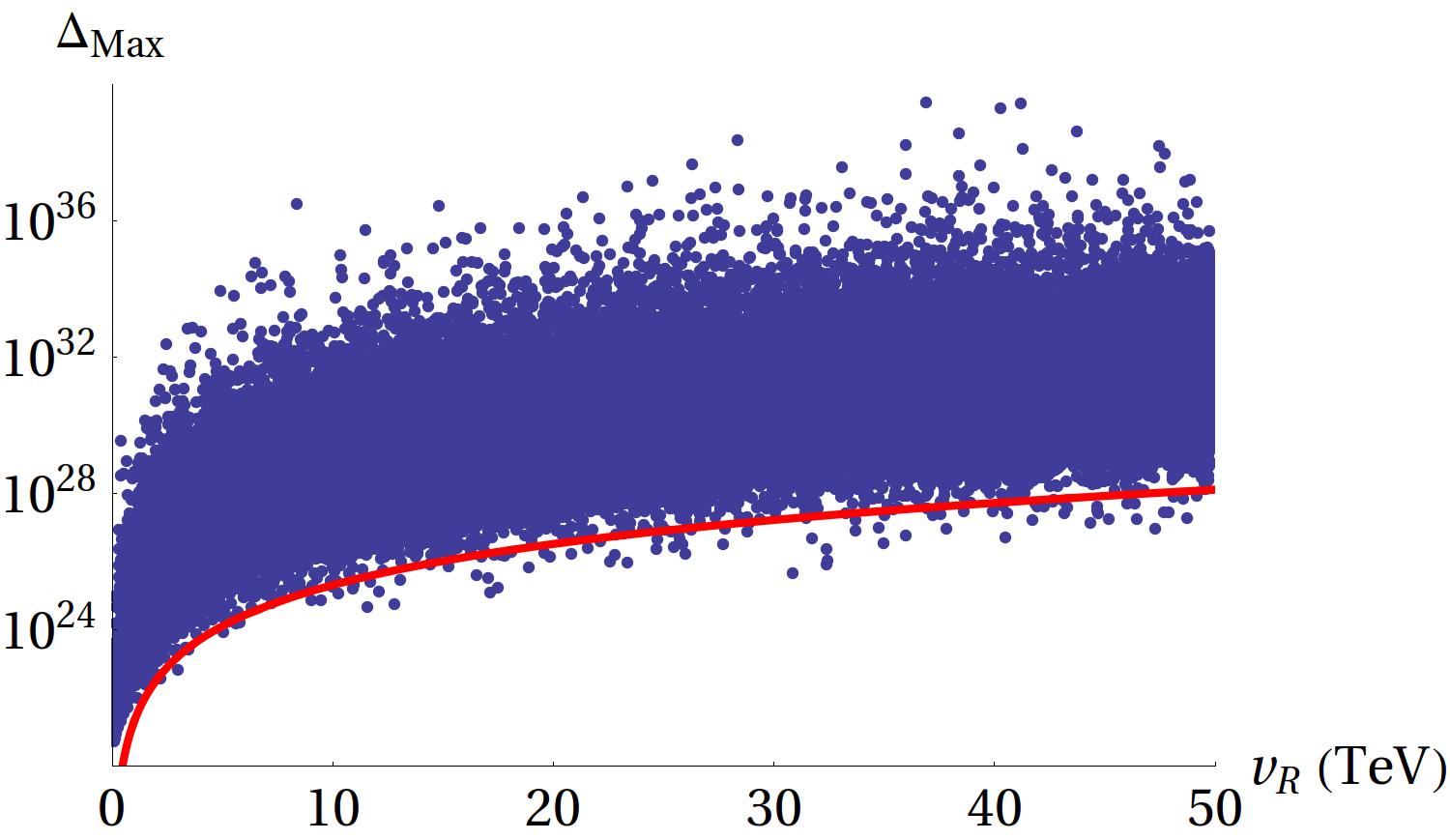} 
\caption{The figure shows the fine-tuning measure $\Delta_{\text{Max}}$ as a function of $v_R$ in TeV for a $CP$-symmetric $V_H$. The red line is parametrized by $1 \cdot 10^{-2} v_R^4/\ka_+\sq v_L\sq$ taking an average value for $v_L$ of $10$ eV.} 
\label{CPfinetune}
\end{figure} 
The Higgs potential in this case is that of Eq.~\eqref{HiggspotCP}. This potential contains $5$ more parameters than the $P$-symmetric potential. Note, however, that if we were to neglect $\tr(\Delta_L \Delta_L^\dagger)$ compared to $\tr(\Delta_R \Delta_R^\dagger)$, we would obtain a  potential very similar to the $P$-symmetric case. Indeed, five of the minimum equations are, up to terms of $\Or(v_L/v_R)$ given by those of the $CP_1$-symmetric case, Eqs.~\eqref{musol}, \eqref{minEq5} and \eqref{minEq6}, with the translations $\mu_3\sq\rightarrow \mu_{3R}\sq$, $\rho_1\sq\rightarrow \rho_{1R}\sq$ and $\al_i\sq\rightarrow \al_{iR}\sq$. The final minimum equation, the vev see-saw relation, is also given by the corresponding $CP_1$ equation, Eq.~\eqref{minEq4}, to $\Or(\ka_+\sq/v_R\sq) $, where $\mu_{3L}\sq/v_R\sq$ now plays the role of $\rho_1$. Schematically,
\bea
2\frac{\mu_{3L}\sq}{v_R\sq}-\rho_3 \sim \frac{\ka_+\sq }{v_L v_R}\bt_i. 
\eea
Thus, unless the $\bt_i$ terms cancel to good precision, the natural value for $\mu_{3L}\sq$ is of the order of $\Or(\frac{v_R}{v_L}\ka_+\sq)$. The $\mu_{3L}\sq$ parameter thereby is the main difference between this and the $CP_1$ case. Since, if $\mu_{3L}\sq= \mu_{3R}\sq  $ we could have identified it with $\mu_3\sq$ of the $CP_1$ case.
Note, however, that $\mu_{3L}$ only appears in the mass terms for the left-handed triplet fields. Furthermore, the two mechanisms which led to small masses of additional Higgs fields in the $CP_1$ case (see section \ref{secCP1}) are still in place here. The equivalent of  Eq.~\eqref{minEq5} now implies $\al_{3R}\sin\al$ to be small. Choosing $\al_{3R}$ small again implies small flavor-changing Higgs masses. On the other hand, a small value of $\al$ implies, through Eqs.~\eqref{minEq6} and \eqref{minEq4}, small $2\mu_{3L}\sq -v_R\sq\rho_3$ which dictates the masses of the left-handed triplet fields. 

Thus, the condition that $\al_{3R}\sin\al$ be small and the lower bound on the flavor-changing Higgs masses force $\al$ to be small. As this is the only source of $CP$ violation in the quark sector, one might expect the model to predict hardly any $CP$ violation. This would lead one to doubt the viability of the model. However, the lower bounds on the $CP$-violating Higgs mass in the $C$ and $P$ cases assumed the relation between the CKM matrices these symmetries imply. 
As there is no such relation in this case the bounds can be weakened. For example, the analysis of Ref.\ \cite{Blanke:2011ry} shows that points in parameters space for values of $M_H$ as low as $M_H\sim 2.4 \,\text{TeV}$ are allowed. Such values still imply a very small $\al=\Or(v_L/v_R)$. However, even smaller values of $M_H$ might be achieved at the price of additional fine-tuning \cite{Blanke:2011ry}, which would allow for larger $\al$.

\subsubsection{Fine-tuning}
\begin{figure}[t]
\centering
$v_L=\bt_i=0$ \hspace*{55mm} $v_L=0,\,\bt_i\neq 0$
\\
\includegraphics[width=75mm]{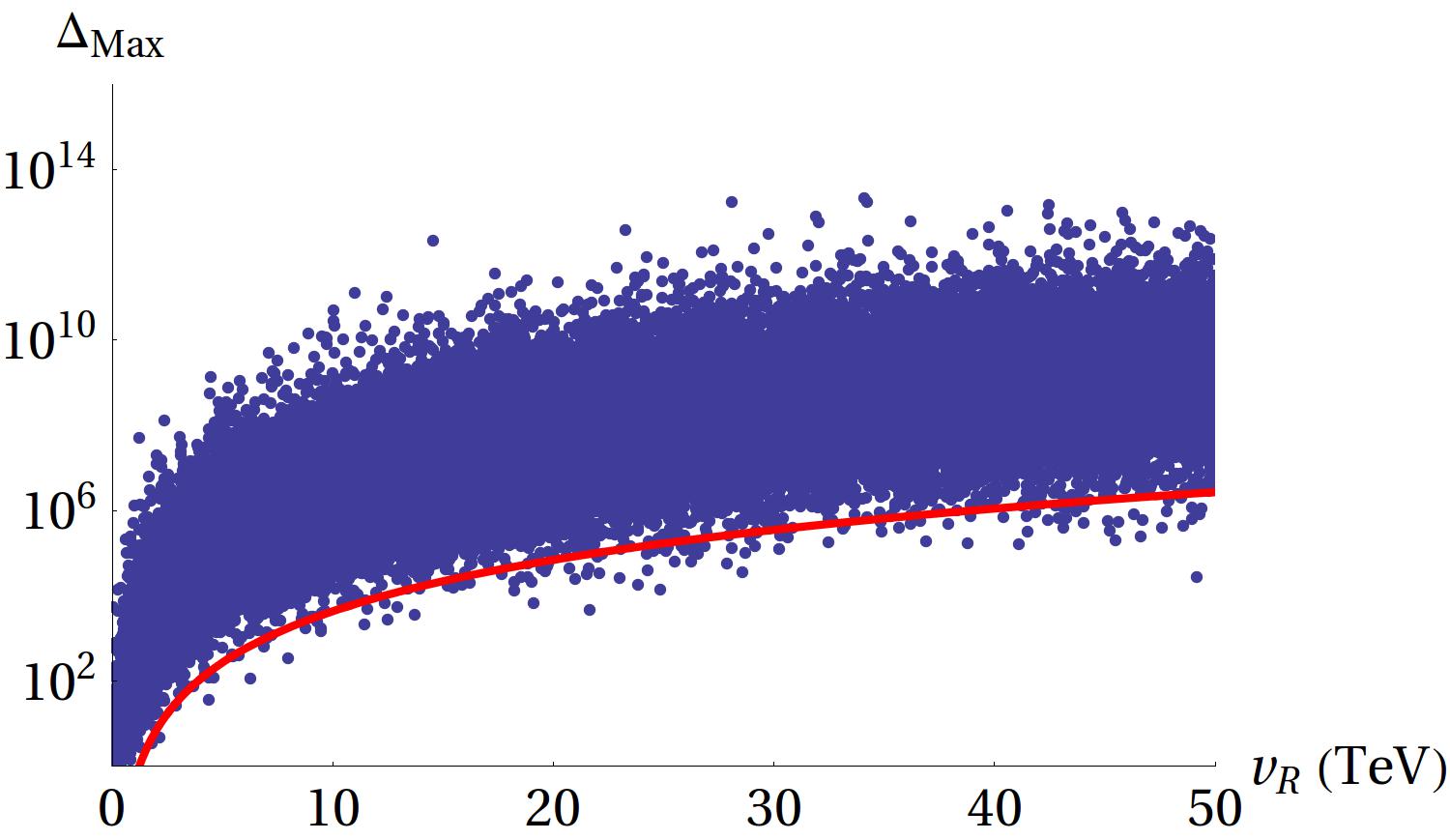} 
\includegraphics[width=75mm]{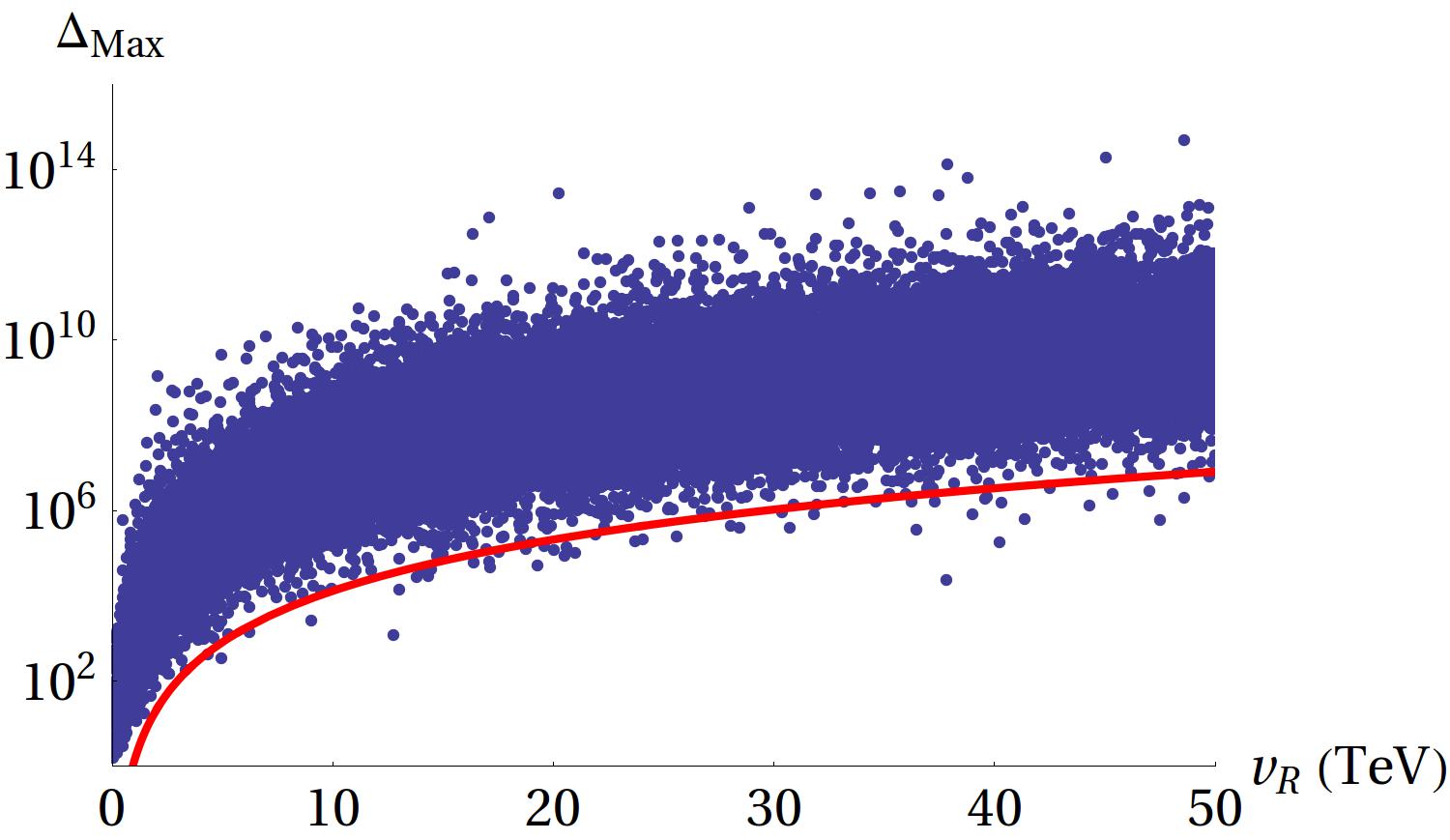} 
\caption{Similar plots to that of Fig.~\ref{CPfinetune}. The plot on the left shows the fine-tuning in the case where $\bt_i$ and $v_L$ are set to zero, while the figure on the right does the same in the case where only $v_L$ is set to zero. The red lines are again chosen such that $0.1\%$ of the points are found below it. It is parametrized by $2\cdot 10^{-3} v_R^4/\ka_+^4$
and $5 \cdot 10^{-3} v_R^4/\ka_+^4$ in the left and right plots, respectively.} 
\label{CPfinetunevL=0}
\end{figure} 
The parameters multiplying the left-handed triplet terms do become important when discussing the fine-tuning in this potential. 
These terms contribute terms to the minimum equations which are smaller than those encountered in the $P$- and $C$-symmetric cases. This means these equations now relate high scales to even smaller scales, indicating more fine-tuning. 

We again go through the same procedure as in the $P$- and $C$- symmetric cases, generating random points in parameter space and calculating the measure of fine-tuning. The results are shown in Figs.~\ref{CPfinetune} and \ref{CPfinetunevL=0}. Clearly, when $v_L\neq0$ and $\bt_i\neq 0$ the fine-tuning measure reaches new heights, due to the vev see-saw relation and the newly added $\tr (\Delta_L \Delta_L^\dagger)\sq$ terms. However, when we choose to eliminate the vev see-saw relation we obtain fine-tuning  measures comparable to the $C$- and $P$-symmetric cases. This may have been expected as when $v_L=0$ the terms which differ from the $P$ case do not contribute to the minimum equations. We then obtain the minimum equations of the $P$-symmetric potential (with $\dt_2=0$), with the translations, $\mu_3\sq\rightarrow \mu_{3R}\sq$, $\rho_1\sq\rightarrow \rho_{1R}\sq$ and $\al_i\sq\rightarrow \al_{iR}\sq$. Thus, as far as the fine-tuning is concerned, when $v_L=0$ the $CP$-symmetric potential simplifies to a special case of the $P$-symmetric potential. As such, the relations between the $\bt_i$ parameters one then obtains is that of the $P$-symmetric case, Eq.\ \eqref{betaRel}.

\section{Summary and conclusions}\label{conclusion}
The most symmetric minimal LR models, the ones invariant under $P$, $C$, and $CP$, turn out not to be viable. There are just two possible implementations of both $P$ and $C$ that are able to produce the observed quark masses, yielding the $CP_1$ and $CP_2$ models. The 
models differ in the relation among the left and right CKM matrices. In the case of the $CP_1$ model this relation puts constraints on $CP$-violating observables from Kaon and $B$-meson mixing that are incompatible with measurements, in particular, the bounds on the $B$-mixing angle $\phi_d$. 
As the Yukawa couplings of the minimal pseudomanifest LRM coincide with those of the $CP_1$ model, it follows that it is also excluded.
The $CP_2$ model cannot be excluded in the same way, as there is in general no simple relation between left and right CKM matrices. In this case 
the Higgs potential is more constraining. Here it was shown that the Higgs potentials of the $CP_2$ model is, to good approximation, equal to a special case of the $CP_1$ invariant potential. For that potential it was shown in Ref.\ \cite{Barenboim:2001vu} that whenever there is spontaneous $CP$ violation, $\al \neq 0$, the potential can not reproduce the SM Higgs spectrum and since the model has no explicit $CP$ violation, the case without spontaneous $CP$ violation has no $CP$ violation at all. Upon field redefinitions, these conclusions carry over to the $CP_2$ model, which can therefore be considered excluded. In addition, both models generally require a large amount of fine-tuning. The minimum equations generally relate very different scales, $\ka_+$, $v_R$ and $v_L$, which implies that some of the parameters will have to be fine-tuned. The most extreme tuning results from the so-called vev see-saw relation Eq.\ \eqref{vevseesaw} \cite{Deshpande:1990ip}, which implies a huge amount of fine-tuning. 

Less symmetric possibilities are the $P$-, $C$- and $CP$-symmetric LRMs, the first two are LR symmetric, while the last possibility, $CP$, does not relate left- and right-handed fields.  The most widely studied case is the $P$-symmetric LRM. This type of LRSM is most constrained in the limit that the ratio of vevs $r\equiv \ka'/\ka$ is small $\lesssim m_b/m_t$. Based on an analytical solution for the phases in the CKM matrices, which allows for strong constraints from $CP$-violating observables Kaon and $B$-meson mixing and the neutron EDM, a lower bound $M_2\gtrsim 10 \, \text{TeV}$ was obtained \cite{Zhang2008}. Outside this regime this solution for the phases does not exist and the bound is weakened, $M_2\gtrsim 3 \, \text{TeV}$ \cite{Bertolini:2014sua}. Although use of a more general solution, recently derived \cite{Senjanovic:2014pva}, may strengthen this bound. This shows that the indirect constraints for $P$-symmetric LRMs are more stringent than the direct limits on $M_2$ for left-right symmetric models, which currently are 2 TeV if one does not wish to make assumptions about right-handed neutrinos. The increase of experimental sensitivity in the coming $10$ years is expected to push the lower indirect bound on $M_2$ to roughly $8$ TeV, thereby exploring a considerable part of the still available parameter space of the $P$-symmetric LRMs. 
The minimal manifest LRSM, the case of vanishing $\al$, is more constrained, here the current bound on the $W_R^\pm$ mass is $M_2\gtrsim 20\, \text{TeV}$ \cite{Maiezza:2014ala}. 

The Higgs potential of the $P$-symmetric LRM has been widely studied in the literature \cite{Gunion:1989in,Basecq:1985sx,Deshpande:1990ip, Barenboim:2001vu}. This potential is very similar to that of the $CP_1$-symmetric LRM \cite{Barenboim:2001vu}, which implies that a SM-like Higgs spectrum is only possible for a specific value of the spontaneous phase. In this case, however, the spontaneous phase can be nonzero, although it is now entirely dictated by the explicit $CP$ violation present in the potential. In a sense this means that the $CP$-violating phase $\al$ is put in by hand and can be as large as allowed by the value of $r$ and the constraints from Kaon and $B$-meson mixing and the nEDM. 

Finally, there is the issue of fine-tuning resulting from the minimum equations. Again the most extreme tuning results from the vev see-saw relation, which implies a huge amount of fine-tuning. It is possible to avoid this by setting $v_L$ and $\bt_i$ to zero by hand as was done in Ref.\ \cite{Zhang2008}. As we have demonstrated, the same reduction in the amount of fine-tuning can be achieved with $v_L=0$ only, in which case there are two relations among the $\beta_i$ parameters \cite{Barenboim:2001vu}. The minimum amount of fine-tuning, assuming $\Or(1)$ parameters, is then found to be $\Delta\sim 100$, where $\Delta$ is a measure for fine-tuning often employed in the study of supersymmetric extensions of the SM, defined in Eq.\ \eqref{FTdef}. It indicates that a change in one parameter by a factor of $\Or(1)$ implies a change in another parameter by a factor $\Or(100)$. This may be acceptable for a theory with such widely varying scales.  

Accepting such a minimum amount of fine-tuning, the $P$-symmetric LRM with some constraints on its $\beta_i$ parameters is still viable, but is expected to be much further constrained in the coming decade. As the lower bound on $M_2$ and hence on $v_R$ increases, the amount of fine-tuning will also increase.   

The $C$-symmetric LRM has been studied less in the literature. In this case the phases in the CKM matrices are free parameters, which can be tuned to avoid constraints from $CP$-violating observables. Nonetheless, a lower-bound similar to the $P$-symmetric case can be set, $M_2\gtrsim 3 \, \text{TeV}$ \cite{Bertolini:2014sua}. 

For the $C$-symmetric and $P$-symmetric LRMs the Kaon and $B$ observables are the most sensitive probes of the mass of the $W_R^\pm$ boson and the most promising way to explore the parameter space of these models. This is due to the fact that these observables are determined by the left- and right-handed CKM elements which, in the $P$/$C$-symmetric case, are constrained by the LR symmetries. Since the relation between the left and right CKM matrices is different for the $C$- and the $P$-symmetric LRMs, these two types of models predict a different pattern in $B_{d,s}$-mixing. Thus, if signs of an LRM are observed in this sector one should in principle be able to tell the difference between $C$- and $P$-symmetric LR models using $B_{d,s}$-mixing observables. EDMs constrain the spontaneous $CP$-violating phase $\al$ and the $W_L-W_R$ mixing angle $\zeta$ and are thereby complementary to the meson-mixing observables. At present the neutron EDM sets a strong limit on these parameters, although the deuteron EDM is expected to be a more sensitive probe by about an order of magnitude. Additionally, in an LR model one expects certain relations to hold between the EDMs of light nuclei, measurements of which would allow for another type of test of LRMs.

Although new phases arise in the Higgs sector of the $C$-symmetric LRM, the Higgs potential is again very similar to that of the $CP_1$-symmetric LRM. A SM-like Higgs spectrum is only possible for a specific value of the spontaneous phase, which is again entirely dictated by the explicit $CP$ violation present in the potential. Perhaps unsurprisingly, the amount of fine-tuning required in the potential is similar to the $P$-symmetric case. By setting $v_L=0$ (and possibly $\bt_i=0$) the vev see-saw relation can be eliminated and the fine-tuning dramatically reduced. Nevertheless, in this case a minimum fine-tuning of $\Delta\sim 100$ is required as well.

Finally, the $CP$-symmetric model, which need not be $P$ and $C$ symmetric separately, is considerably different from the above cases. This is due to the fact that $CP$ is not a LR symmetry. Although, like the $P$ case, there is a bound on $r\sin\al$, the CKM matrices and coupling constants are now generally unrelated to one another ($g_L\neq g_R$). This means there is more freedom in this model, such that {\it both} direct and indirect bounds are expected to be weakened, see, for instance, Ref.\ \cite{Blanke:2011ry}. 
The relation between the EDMs of light nuclei is unaffected by this and would still allow for a test of the model. 
As most features of the Higgs sector again resemble the $CP_1$ case, a nonzero spontaneous phase again requires light Higgs fields. However, due to the additional freedom, the bounds on these fields may be weaker in this case. On the other hand, the potential generally requires more fine-tuning than the $CP_1$ potential, except when considering the $v_L=0$ cases, in which the fine-tuning is similar to the $P$-symmetric case. 

Recapitulating, LRSMs are possibly the most attractive of the possible LRMs, but also the most constrained. The LR scale of these models is currently constrained to be in the TeV range, $M_2\gtrsim 3\, \text{TeV}$. Instead, $CP$-symmetric LRMs seem less constrained, allowing for more freedom in the right-handed CKM matrix. In the Higgs sector the potentials of the LRSMs all turn out to be quite similar. The $CP$-symmetric case allows some more freedom in the masses of the left-handed triplet fields, but is otherwise similar as well. The Higgs potentials of all three models require a  considerable amount of fine-tuning, which poses the biggest challenge to their viability. 

\section*{Acknowledgments}
We would like to thank Wilco den Dunnen, Robert Fleischer, Jean-Marie Fr\`ere and Jordy de Vries for useful discussions.  

\bibliography{LRrefs}
\bibliographystyle{h-elsevier} 
\end{document}